\documentclass[useAMS,usenatbib,referee]{mn2e}

\usepackage{multirow}
\usepackage{graphicx,color}
\newcommand{\etal }{{et al.} }
\newcommand{\msun}{\thinspace M_\odot}

\def\lesssim{\mathrel{\hbox{\rlap{\hbox{\lower4pt\hbox{$\sim$}}}\hbox{$<$}}}}
\def\gtrsim{\mathrel{\hbox{\rlap{\hbox{\lower4pt\hbox{$\sim$}}}\hbox{$>$}}}}
\newcommand{\cm}{\,{\rm cm}^{-3} }

\newcommand{\dfrac}[2]{{\displaystyle \frac{#1}{#2}} }

\title[Circumstellar Disk Formation]{Conditions for Circumstellar Disk Formation: Effects of Initial Conditions and Sink Treatment}

\author[M. N. ~Machida, \etal]
  { Masahiro N. Machida$^{1}$\thanks{E-mail: machida.masahiro.018@m.kyushu-u.ac.jp (MNM)}, Shu-ichiro Inutsuka$^{2}$ and Tomoaki Matsumoto$^3$\\
$^{1}$ Department of Earth and Planetary Sciences, Faculty of Sciences, Kyushu University, Fukuoka 812-8581, Japan\\
$^{2}$ Department of Physics, Graduate School of Science, Kyoto University, Sakyo-ku, Kyoto 606-8502, Japan\\
$^{3}$ Faculty of Humanity and Environment, Hosei University, Fujimi, Chiyoda-ku, Tokyo 102-8160, Japan
}

\begin{document}

\maketitle

\begin{abstract}
The formation of a circumstellar disk in collapsing cloud cores is investigated  with three-dimensional magnetohydrodynamic simulations.
We prepare four types of initial cloud having different density profiles and calculate their evolution with or without a sink.
To investigate the effect of magnetic dissipation on disk formation, the Ohmic dissipation 
is considered in some models.
Calculations show that disk formation is very sensitive to both the initial cloud configuration and the sink treatment.
The disk size considerably differs in clouds with different density profiles even when the initial clouds have almost the same mass-to-flux ratio.
Only a very small disk ($\sim10$\,AU in size) appears in clouds with a uniform density profile, whereas a large disk ($\sim100$\,AU in size) forms in clouds with a Bonnor-Ebert density profile.
In addition, a large sink accretion radius numerically impedes disk formation during the main accretion phase and tends to foster the misleading notion that disk formation is completely suppressed by  magnetic braking.
The protostellar outflow is also greatly affected by the sink properties. 
A sink accretion radius of $\lesssim 1$\,AU and sink threshold density of $\gtrsim 10^{13}\cm$ are necessary for investigating disk formation during the main accretion phase.

\end{abstract}
\begin{keywords}
accretion, accretion disks---ISM: jets and outflows, magnetic fields---MHD---stars: formation, low-mass
\end{keywords}

\section{Introduction}
\label{sec:intro}
The circumstellar disk plays a crucial role in star and planet formation.
Gas in the infalling envelope is thought to initially accrete onto the circumstellar disk.
Part of the gas in the disk is ejected by the protostellar outflow or stellar wind and returns to interstellar space, whereas the remainder finally falls onto the protostar and contributes to protostellar mass growth.
In the circumstellar disk, the solid component settles into the disk equatorial plane, and rocky cores or rocky planets form \citep{hayashi85}.
Then, a rocky core acquires its mass from the circumstellar disk and evolves into a gas giant planet \citep{mizuno78,mizuno80}.
Thus, without considering the circumstellar disk, we cannot investigate the formation of stars and planets.

Despite its importance, the formation process of the circumstellar disk is still controversial. 
Observationally, circumstellar disks were first identified by their spectral energy distribution \citep[e.g.][]{lada87}. 
Recent direct observations have yielded various types of information on circumstellar disks \citep{watson07}.
The observations indicate that a sufficiently mature  disk generally exists around Class I and II protostars \citep{andrews05,williams11}.
On the other hand, the circumstellar disk is considered to grow during the Class 0 phase, because a dense infalling envelope can provide sufficient gas to the disk \citep{andre93}.
Therefore, we have to observe circumstellar disks around Class 0 protostars to investigate disk formation; however a dense infalling envelope impedes direct observation of these disks.
Although very recent observations confirm the existence of a (forming or growing) circumstellar disk around class 0 protostars \citep[e.g.][]{jorgensen07, enoch09, enoch11, tobin12}, a theoretical approach is required to clarify the formation and evolution of the circumstellar disk.

Because  a prestellar cloud core has an angular momentum much larger than that of the star \citep{goodman93,caselli02}, the formation of the circumstellar disk has been considered a natural consequence of angular momentum conservation.
However, in addition to the angular momentum or rotation, the magnetic field is also an important ingredient in circumstellar disk formation.
The angular momentum in a star-forming core can be transferred by magnetic effects such as  magnetic braking \citep{shu87,basu94,galli06} and protostellar outflow \citep{tomisaka98,tomisaka02}. 
The magnetic field can be amplified by rotation or shearing motion around the protostar, and  an amplified field effectively transfers the angular momentum outward, slowing the disk rotation.
Therefore,  it is very  difficult to analytically investigate disk formation in the collapsing cloud considering the effects of both the rotation and magnetic field.
As a result, it is necessary to use  numerical simulations  to investigate the formation and evolution of the circumstellar disk.

For this, we need to calculate the evolution of the collapsing cloud (or molecular cloud core) from the prestellar stage while resolving the region where the circumstellar disk forms.
However, the spatial scales and timescales of the molecular cloud core and circumstellar disk considerably differ.
The difference between the spatial scales of the (prestellar) molecular cloud core ($\sim10^5$\,AU) and circumstellar disk ($\sim1-10$\,AU) or protostar ($\lesssim 0.01$\,AU) has been overcome with adaptive mesh refinement \citep[e.g.][]{hennebelle09, joos12, seifried12} or nested grid \citep[e.g.][]{machida11b} techniques. 
However, even with such techniques, the different timescales of the protostar ($\sim$\,days) and the collapsing cloud ($\sim10^5$\,yr) make it difficult to calculate the cloud evolution for the long period until a mature disk forms.
To resolve this issue, the sink cell approach is usually adopted.
With this approach, the region around a protostar is masked by sink cells, and a high-density gas, which has a shorter dynamical timescale, is removed from the computational domain and  added to the protostellar mass.
Although we do not know  whether the sink cell treatment gives the correct answer for disk formation, we have tentatively investigated  disk formation with it.


Recent studies of disk formation in the collapsing cloud pointed out that the excessive angular momentum near the protostar is transferred by magnetic braking, and disk formation is strongly suppressed, at least during the early phase of star formation. 
Using two-dimensional ideal magnetohydrodynamic (MHD) calculations with sink a radius of $r_{\rm acc}=6.7$\,AU, \citet{mellon08} showed that disk formation is suppressed by magnetic braking even when the initial cloud is weakly magnetized with  $\mu \lesssim 10$, where $\mu$ is the dimensionless mass-to-flux ratio in units of the critical value $(2 \pi G^{1/2})^{-1}$ and is frequently used as an indicator of cloud magnetisation. 
Using three-dimensional ideal MHD calculations, \citet{seifried12} investigated high-mass star formation in a massive cloud and showed that the circumstellar disk forms only with $\mu \gtrsim 10$; they adopted a the sink accretion radius of $r_{\rm acc}=12.6$\,AU and sink threshold density of $\rho_{\rm thr} = 1.78\times10^{-12}$\,g\,$\cm$ (or $n_{\rm thr}=5\times10^{11}\cm$).
However, these studies may overestimate the effects of the magnetic field or the efficiency of magnetic braking, because in reality, the magnetic field dissipates in the high-density gas region \citep[e.g.][]{nakano02,kunz10}.
Considering  magnetic dissipation by ambipolar diffusion and Ohmic dissipation (and the Hall effect), \citet{li11} investigated disk formation using two-dimensional calculations with a sink radius of  $r_{\rm acc}=6.7$\,AU (see also \citealt{mellon09,krasnopolsky10, krasnopolsky11} ).
They claimed that magnetic dissipation cannot drastically assist disk formation, and thus a rotation-supported disk never forms with $\mu \lesssim 10$.

These studies suggested that magnetic braking severely suppresses disk formation in the early phase of star formation.
However, only a massive infalling envelope can brake the circumstellar disk, and the angular momentum transferred by magnetic braking is stored  in the infalling envelope. 
Thus, magnetic braking gradually becomes ineffective as the infalling envelope dissipates.
In addition, the infalling envelope finally falls onto the centre of the cloud, and the falling gas brings the angular momentum into the circumstellar region.
Therefore, a rotation supported-disk is expected to appear as the mass of the infalling envelope decreases.
Using three-dimensional resistive MHD calculations with a sink radius of $r_{\rm acc}=1$\,AU and threshold density of $n_{\rm thr}=10^{13}\cm$, \citet{machida11b} investigated the evolution of a cloud with a finite envelope mass and showed that a sufficiently large rotation-supported disk can form in the later main accretion phase even when the initial cloud is strongly magnetized with $\mu \sim1-3$. 
In addition to the effect of the infalling envelope, the initial configuration of the magnetic field lines and rotation axis may affect disk formation.
\citet{li11}, \citet{machida11b} and \citet{seifried12} assumed the initial magnetic field lines parallel to the initial rotation axis.  
On the other hand, \citet{hennebelle09} and \citet{joos12} adopted magnetic field lines not parallel to the cloud rotation axis as the initial state.
They calculated the evolution of such a cloud without a sink but with a spatial resolution of $\sim0.5$\,AU and found that a circumstellar disk can form in the cloud  with $\mu \lesssim3 $ even without magnetic dissipation (i.e. the ideal MHD limit).

At the expense of the spatial resolution around the protostar, many studies realized the long-term evolution of a collapsing cloud with a relatively large sink radius ($r_{\rm acc} \gtrsim1$\,AU or a coarser spatial resolution of $\gtrsim0.1$\,AU). 
On the other hand, \citet{dapp10} and \citet{dapp12} investigated disk formation with a much  smaller sink radius of $\sim 0.02$\,AU with two-dimensional simulations including magnetic dissipation.
They showed  that a low-mass disk with a radius of $0.05$\,AU forms just after protostar formation because  magnetic dissipation reduces magnetic braking around the protostar. 
They also pointed out that a disk $\sim10$\,AU in size remains during the main accretion phase even in a strongly magnetized cloud of $\mu = 2$ \citep[see also][]{machida11c}.

It seems that each study reported different outcomes of disk formation.
For example, \citet{li11} claimed that a circumstellar disk forms only in a very weakly magnetized cloud of $\mu > 10$, whereas \citet{dapp12} showed that a rotation-supported disk forms even in a cloud with $\mu \sim 2$.
These different results may be caused by different initial settings and sink treatment.
In addition, different cloud parameters such as the cloud mass, radius and magnetic field distribution should affect disk formation.
In this study, we investigate the effect of the initial conditions and sink treatment on disk formation, by calculating 24 models with different initial conditions and sink properties.
To limit the number of calculation models, we adopt only Ohmic dissipation as the magnetic dissipation process.
Ohmic dissipation becomes effective only in a high-density gas region.
Thus, our calculation imposes very severe limitations on  the disk formation because other processes (ambipolar diffusion and the Hall effect) support disk formation in a lower-density gas region.

This paper is structured as follows. 
The numerical settings and initial conditions are described in \S2. 
The calculation results are given in \S3.
We discuss the effect of the initial conditions and sink treatment on disk formation in \S4 and summarize the disk formation process in \S5.

\section{Numerical Settings and Initial Conditions}
\subsection{Numerical Method}
\label{sec:model}
Starting from a prestellar cloud core, we calculate the cloud evolution for a longer duration after  protostar formation to investigate the formation of a circumstellar disk during the main accretion phase.
To cover the very different spatial scales of the prestellar cloud core ($\sim10^3-10^5$\,AU) and circumstellar disk ($\sim1-100$\,AU), we use a nested grid composed of rectangular grids with the same number of cells (i, j, k) = (64, 64, 32), in which mirror symmetry with respect to the $z=0$ plane is imposed (for the details of the nested grid method, see \citealt{machida05a,machida05b}).

We construct four different initial clouds having different density distributions (for details, see \S\ref{sec:initial}) and calculate the evolution of each cloud with the (resistive) MHD equations (see eqs. [1] - [5] of \citealt{machida11b}).
In each calculation, we prepare 5 -- 9 grid levels at the beginning;  the initial cloud is immersed in the fifth level in all the models.
The grid size and cell width are halved with each increment in the grid level $l$.
To reduce the effect of the reflection of Alfv$\acute{e}$n waves at the boundary, we impose the computational boundary at about 6 -- 16 times  the initial cloud radius \citep{machida06b,machida11a,machida11b,machida12}.
Thus, the $l=1$ grid  has a box size of $L(1) = 12 - 32\, r_{\rm c}$, where $r_{\rm c}$ is the initial cloud radius. 
A low-density interstellar gas with a density of $10^3\cm$ is set outside the initial cloud.

We impose gravity (both gas self-gravity and protostellar gravity) only inside the star forming cloud core (i.e. the region of $r<r_{\rm c}$).
Thus, only the gas at  $r<r_{\rm c}$ collapses toward the centre of the cloud.
After the calculation starts, a new finer grid is dynamically generated at the centre of the collapsing cloud, and the Truelove condition \citep{truelove97} is satisfied, in which the Jeans length is resolved in at least eight cells.
The maximum grid level is restricted to $l=11-18$, which have a spatial resolution of 0.06 -- 4.7\,AU (for details, see \S\ref{sec:results}). 
In most calculations, sink cells are introduced and the high-density gas is removed from the computational domain (for details, see \S\ref{sec:sink}).
Because the maximum grid level and the physical size of the first grid level differ for each model, we describe them in \S\ref{sec:results} in more detail.

For all calculation models, instead of solving the energy equation, we use a barotropic equation of state, 
\begin{equation} 
P =  c_{s,0}^2\, \rho \left[ 1+ \left(\dfrac{\rho}{\rho_c}\right)^{2/5} \right],
\label{eq:eos}
\end{equation}
where  $ \rho_c = 10^{-13} \, \rm{g} \, \cm$ ($n_c \simeq 2.5\times 10^{10} \cm$) is adopted as the critical density.
The critical density is almost identical to that in previous studies \citep{mellon08,mellon09,li11,machida11b,dapp12,joos12}.
Note that although the critical density slightly differs among previous studies \citep[e.g.][]{li11,machida11b,joos12}, the same value of the critical density is introduced in all the models in order to focus only on the effects of the sink and initial conditions on  disk formation. 
A sound speed of $c_{s,0} \simeq 0.2$\,km\,s$^{-1}$ is adopted in most models (low-mass star formation models), although  $c_{s,0} = 0.3$\,km\,s$^{-1}$ is adopted in some models (massive star formation models).
Equation~(\ref{eq:eos}) indicates that the gas behaves isothermally for $\rho < \rho_c$, whereas it behaves adiabatically for $\rho>\rho_c$.

As the dissipation of magnetic field, we assume only Ohmic dissipation, as described in \S\ref{sec:intro}.
The resistivity $\eta$ is related to the ionisation degree $X_{\rm e}$ \citep{nakano02,machida07,dapp10} and is simply described as
\begin{equation}
\eta = \dfrac{740}{X_e} \sqrt{\dfrac{T}{{10\,{\rm K}}}}\ \ \ {\rm cm}^2\,{\rm s}^{-1}.
\label{eq:eta}
\end{equation}
where $T$ and $n$ are the gas temperature and number density, respectively, and $X_e$ is the ionisation degree of the gas, which defined as
\begin{equation}
X_e =  5.7 \times 10^{-4} \left( \dfrac{n}{{\rm cm}^{-3}} \right)^{-1}.
\label{eq:xe}
\end{equation} 
Because the temperature is a function of the density (eq.[\ref{eq:eos}]), the resistivity is also a function of the density. 
Equations~(\ref{eq:eos}) and (\ref{eq:xe}) are not adequate in the high-density gas region of $n\gtrsim 10^{15}\cm$ \citep{machida07}. 
However, because we calculate the cloud evolution with a sink that limits the maximum density to $n<10^{15}\cm$ (\S\ref{sec:sink}), we can safety ignore the thermal evolution and resistivity in the high-density gas region in this study.
In each model, the mass-to-flux ratio of the initial cloud $\mu$ is numerically calculated as
\begin{equation}
\mu = f_\mu \left( \dfrac{M}{\phi} \right), 
\end{equation}
where $f_\mu = (2 \pi\,G^{1/2})^{-1}$ is used as the critical value of the mass-to-flux ratio. 

\subsection{Initial Conditions}
\label{sec:initial}
To investigate disk formation in clouds with different initial configurations, we prepared the 24 models listed in Table~\ref{table:1}.
The models have four different density distributions of the initial cloud. 
In Table~\ref{table:1}, the first two characters of the model name indicate the density distribution of the initial cloud.
Models US1 -- US8 and USL have a uniform density, and the cloud parameters are identical to those of the fiducial models of \citet{li11} and \citet{krasnopolsky12}.
Models BE1 -- BE6 and BEH have a Bonnor-Ebert density profile and are identical to the fiducial model of \citet{machida11b}.
Models RJ1 -- RJ4 have the same initial density distribution and cloud parameters as those in \citet{hennebelle09} and \citet{joos12}.
A relatively massive initial cloud is adopted for models RS1 - RS4, which have the same density profile and cloud parameters as in \citet{seifried12}.

In all the models, rigid rotation is adopted in the initial cloud core.
In addition, we only assume initial magnetic field lines parallel to the initial rotation axis;  both the rotation axis and magnetic field lines are parallel to the $z$-axis.
As described in \S\ref{sec:model}, the same equation of state and resistivity are used in all the models.
Note that a resistivity of $\eta=0$ is adopted in some ideal MHD models.
The initial number density $n_{\rm c,0}$, cloud mass $M_{\rm c}$, cloud radius $r_{\rm c}$,  magnetic field strength (at the centre) $B_0$ and angular velocity $\Omega_0$ are listed in Table~\ref{table:1}.
An initially uniform magnetic field is adopted in models labeled (U) in the fifth column of Table~\ref{table:1}, whereas a non-uniform magnetic field is adopted in the others.
The ratios of the thermal ($\alpha_0$), rotational ($\beta_0$) and magnetic ($\gamma_0$) energy to the gravitational energy are also listed in Table~\ref{table:1}.

\subsection{Sink Treatment}
\label{sec:sink}
A circumstellar disk forms around a protostar.
Thus, in principle, we should need to resolve the protostar itself to investigate the formation of the circumstellar disk.
However, the calculation timestep becomes increasingly shorter as the spatial resolution improves, and we cannot calculate the evolution of the collapsing cloud until disk formation \citep[e.g.][]{machida11c,dapp10,dapp12}. 
For this reason, a sink cell is often introduced to make long-term calculation possible.
The protostar is masked and treated as a gravitating point source with the sink treatment; thus, we do not resolve the region inside the sink (or accretion) radius.

Usually the sink is described by two parameters: the accretion radius $r_{\rm acc}$ and threshold density $\rho_{\rm thr}$ or $n_{\rm thr}$ \citep{krumholz04, federrath10,machida10a,machida11a,seifried12}.
We identified the protostar formation epoch $t_{c}=0$ as that at which the maximum density, which corresponds to the density at the centre of the collapsing  cloud, reaches $\rho = \rho_{\rm thr}$.
Then, after protostar formation, the mass exceeding the threshold density $\rho > \rho_{\rm thr}$ inside the accretion radius $r<r_{\rm acc}$ is removed from the computational domain and added to the protostellar mass or protostellar gravity.
Thus, for each timestep, the accretion mass $M_{\rm acc}$ onto the protostar is estimated as
\begin{equation}
M_{\rm acc} = \int_{r<r_{\rm acc}} [\rho \, (i, j, k) - \rho_{\rm thr}]\, dV.
\end{equation}
Because we adopt a spherically symmetric or axisymmetric cloud as the initial state, the protostar should be fixed at the centre of the cloud.
Thus, in this study, the sink cells are fixed at the centre of the computational domain.
We may have to use a moving sink with additional sink criteria \citep{federrath10,krumholz04} even when the initial cloud is spherically symmetric because non-axisymmetric perturbation can develop.
However, we use this simple setting for the sink to compare previous studies with this study.
Note that almost the same sink criteria as in this study were used in \citet{machida11a}, \citet{li11}, \citet{seifried12} and \citet{dapp12}.
For further comparison of the calculation without a sink \citep{hennebelle09, joos12}, we do not impose the sink in some models, in which the adiabatic equation of state (eq.~[\ref{eq:eos}]) and limited spatial resolution prevent the cloud from further collapse. 
Whether the sink is introduced (Y) or not (N) is listed in the eleventh  column of Table~\ref{table:1}.

To investigate the effects of the sink on circumstellar disk formation and determine the necessary condition for the sink, we change both the accretion radius $r_{\rm acc}$ and threshold density $n_{\rm thr}$ (or $\rho_{\rm thr}$) in each model.
The accretion radius is listed in the twelfth column of Table~\ref{table:1}.
We adopted two type of threshold density: variable threshold density (V) and fixed threshold density.
In the former, the (maximum) density on the cell just outside the sink cell is used as the threshold density. 
Thus, the threshold density varies during the calculation.
This treatment mimics the outflow boundary condition adopted in the studies of \citet{li11} and his collaborators, in which the density and velocity components are copied from the first active zone into the sink region along the radial direction \citep{mellon08}.
In the latter, the threshold density, which is listed in the thirteenth column of Table~\ref{table:1}, is fixed during the calculation, which corresponds to the sink treatment in \citet{machida11b} and \citet{seifried12}.
The cell width in the finest grid is listed in the fourteenth column in Table~\ref{table:1}.
Whether magnetic dissipation (or Ohmic dissipation) is included (Y) or not (N) is noted in the fifteenth column of Table~\ref{table:1}.

\section{Results}
\label{sec:results}
\subsection{Disk Formation in a Uniform Sphere}
\label{sec:uniform}
\subsubsection{Initial Cloud Properties and Numerical Settings}
For a uniform sphere model, which corresponds to models US1 -- US8 and USL in Table~\ref{table:1}, we adopted the same cloud parameters as in the fiducial model in \citet{li11}. 
Note that the numerical settings and cloud evolution for model USL are described in \S\ref{sec:resolution}.
The cloud has a uniform density $n_{\rm c,0}=10^5\cm$, uniform magnetic field $B_0 = 35.4\,\mu$\,G and  rigid rotation with an angular velocity of $\Omega_0 = 10^{-13}$\,s$^{-1}$.
The mass and radius of the cloud are $M_{\rm c}=1\msun$ and $r_{\rm c}=6.7\times10^3$\,AU, respectively. 
The initial cloud has a mass-to-flux ratio of $\mu=2.9$.
Although the grain size distribution and cosmic ray ionisation rate, which are related to the magnetic dissipation process, were changed in \cite{li11}, a rotation-supported disk never forms with this cloud parameter.
\citet{li11} included both ambipolar diffusion and Ohmic dissipation as the magnetic dissipation processes, whereas we considered only Ohmic dissipation.
Thus, we may underestimate the degree of magnetic dissipation and overestimate the magnetic field strength and efficiency of magnetic braking near the centre of the cloud to some degree.

The initial cloud is immersed  in the fifth level of the grid, which has a box size of $L(5)=1.34\times10^4$\,AU ($=2\,r_{\rm c}$).
The first level of the grid has a box size of $L(1) = 2.2\times10^5$\,AU.
We prepared nine nested grids ($l$ = 1 -- 9) initially, and the maximum grid level was restricted to $l=14$ (models US1 -- US6), 15 (model US7) and 16 (model US8).
The maximum grid level has a box size of $L(14)=26$\,AU (US1 -- US6), $L(15)=13$\,AU (US7) and $L(16)=7$\,AU (US8), and a cell width of $h(14)=0.4$\,AU (US1 -- US6), $h(15)=0.2$\,AU (US7) and $h(16)=0.1$\,AU (US8).

Models US1 -- US8 have the same cloud parameters listed in Table~\ref{table:1}, but they have different sink properties.
For the sink, only the accretion radius (and variable threshold density) is imposed for models US1 -- US4, whereas both the accretion radius and threshold density are imposed for models US5 -- US8.
Thus, the threshold density is determined in each timestep for models US1 -- US4 (\S\ref{sec:sink}), in which the accretion radius differs among the models.
On the other hand, for models US5 -- US8, a fixed threshold density in the range of $10^{11}\cm\le n_{\rm thr} \le 10^{14}\cm$ is adopted for each model,  but the accretion radius is fixed at $r_{\rm acc}=1$\,AU.
Ohmic dissipation is imposed for all the US models. 
The protostar formation epoch $t_c=0$ is determined as that at which the maximum grid level is generated for models US1 -- US4, whereas it is identified as that at which the maximum (or central) density reaches the threshold density $n_{\rm thr}$ for models US5 -- US8.

\subsubsection{Calculation Results}
\begin{figure}
\includegraphics[width=150mm]{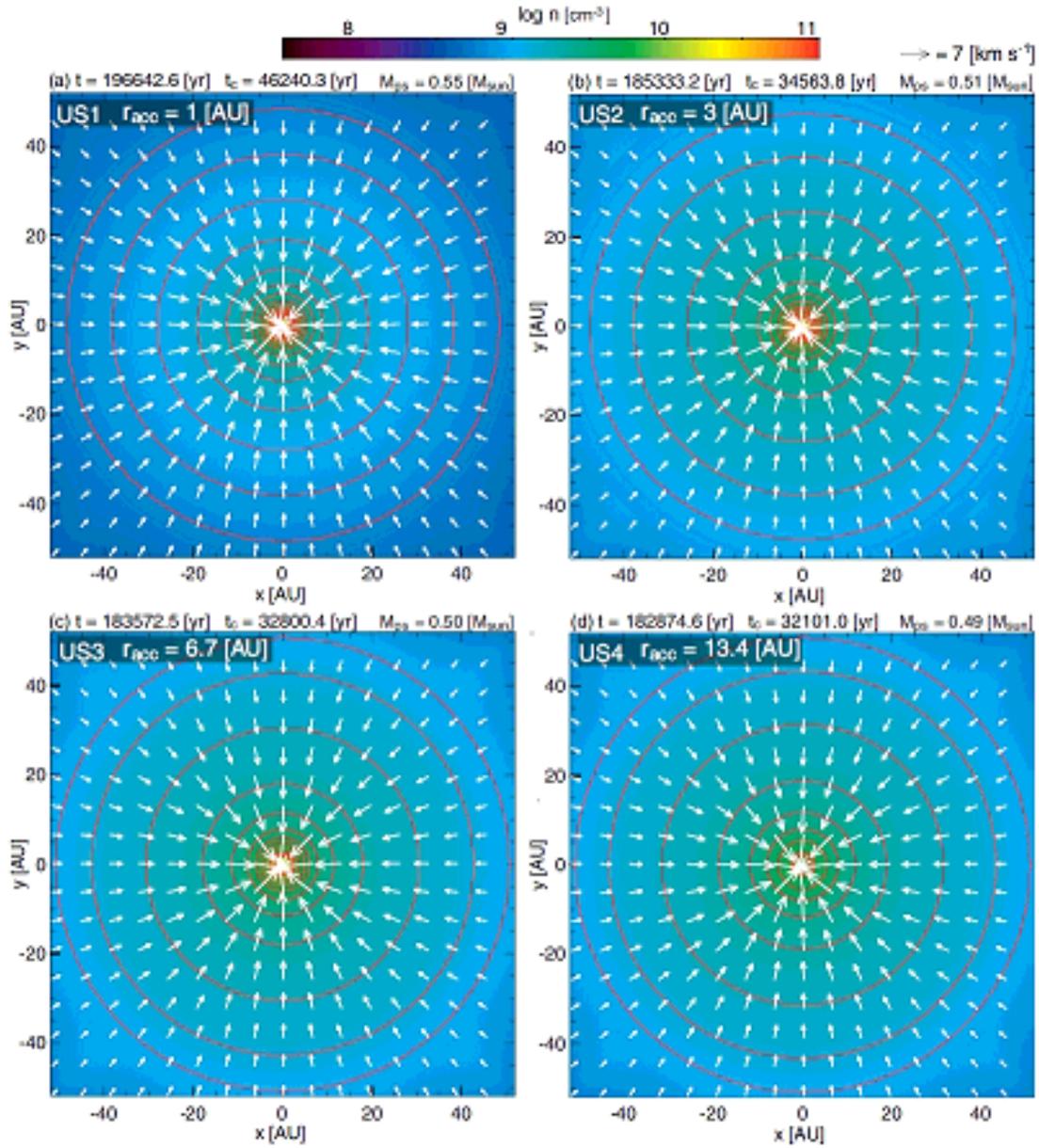}
\caption{
Density (colour and contours) and velocity (arrows) distributions on the equatorial plane for models US1, US2, US3 and US4 when the protostellar mass reaches $M_{\rm ps }\simeq 0.5\msun$.
Elapsed time after the cloud begins to collapse $t$ and that after protostar formation $t_c$ and  protostellar mass $M_{\rm ps}$ are given in each panel.
Model name and sink accretion radius $r_{\rm acc}$ are also given.
}
\label{fig:1}
\end{figure}

\begin{figure}
\includegraphics[width=150mm]{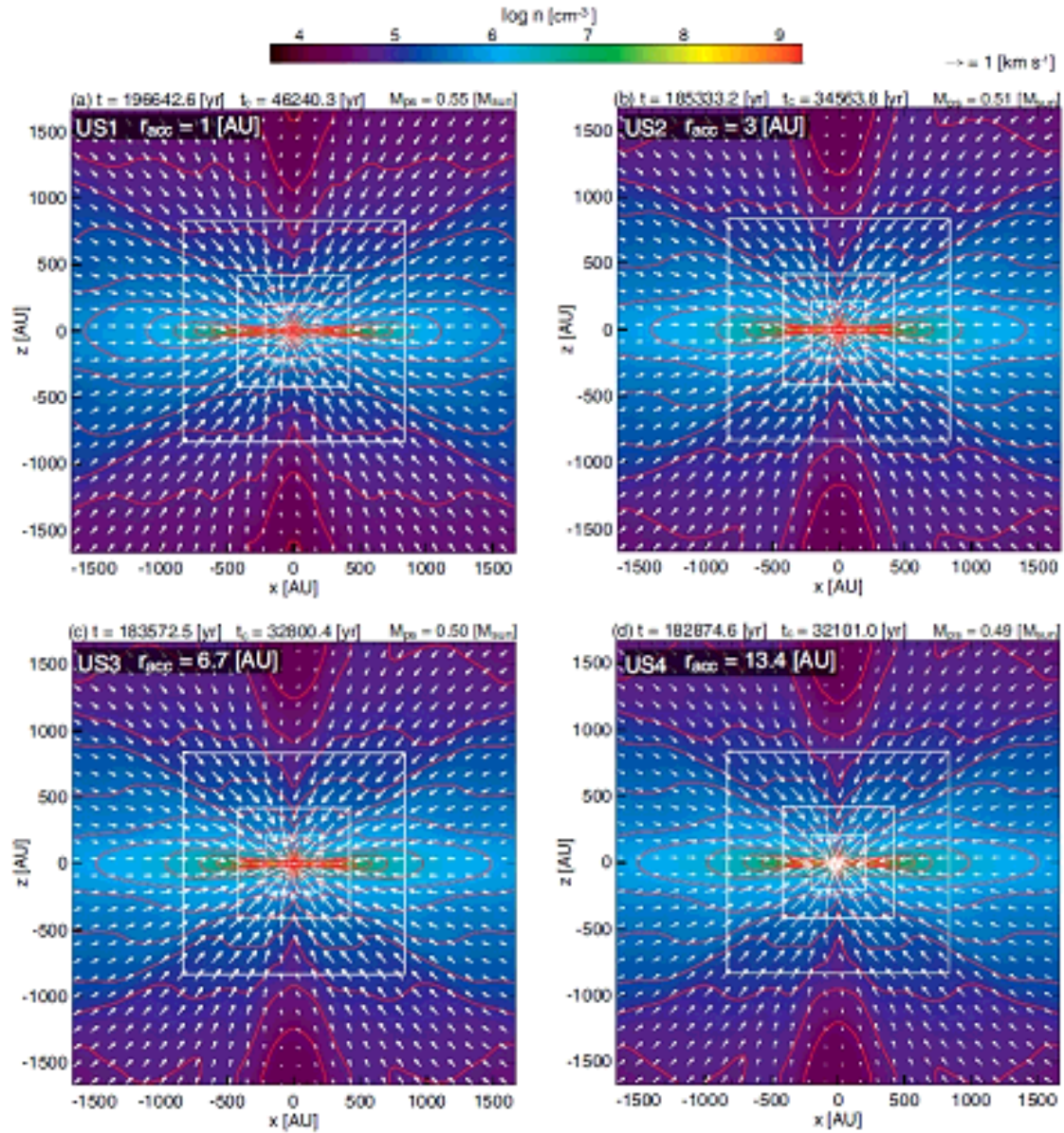}
\caption{
Same as Fig.~\ref{fig:1}, but on the $y=0$ plane.
Box scale differs from that in Fig.~\ref{fig:1}.
}
\label{fig:2}
\end{figure}

Figures~\ref{fig:1} and \ref{fig:2} show the density and velocity distributions on the equatorial (Fig.~\ref{fig:1}) and $y=0$ (Fig.~\ref{fig:2}) planes for models US1, US2, US3 and US4 when the protostellar mass 
reaches $M_{\rm ps }\simeq 0.5\msun$.
Because the initial cloud has a mass of $1\msun$,  half of the cloud mass has fallen onto the protostar by this epoch.
In other words, almost half of the initial cloud mass remains in the infalling envelope.
Thus, by definition, this epoch corresponds to the end of the Class 0 stage or the beginning of the Class I stage \citep{andre93,andre94}.  

In each model, the accretion radius $r_{\rm acc}$ is different, and  the threshold density is determined in each timestep as described in \S\ref{sec:sink}.
Figure~\ref{fig:1} shows that for each model, the gas falls directly to the centre of the cloud without rotation, and the radial velocity $v_r$ strongly dominates the rotation velocity $v_\phi$.
The figure also shows that no rotation disk exists around the protostar at this epoch. 
However, during the early main accretion phase (i.e. $M_{\rm ps} \ll 0.5 \msun$), the cloud evolution differs among the models. 
A disk-like structure appears around the centre of the cloud just after protostar formation in models US1 and US2, whereas it never appears in models US3 and US4.
In models US1 and US2, the disk, which is partly supported by rotation is $\sim5$\,AU in size and persists for $t_c \sim 6\times 10^3$\,yr after protostar formation.
Then, the disk gradually shrinks and completely disappears at $t_c\sim10^4$\,yr. 
After the disappearance, the disk never appears in models US3 and US4.
As a result, although the cloud evolution during the early accretion phase differs qualitatively with different sink treatments, no persistent disk appears for models US1 -- US4.

\cite{larson69} and \cite{masunaga00} showed that an adiabatic core $\sim$\,1\,AU in size (hereafter the first core) appears before  protostar formation.
The hydrodynamical simulations by \cite{bate98} and \citet{machida10a} showed that the first core becomes a circumstellar disk, or rotation supported disk, after protostar formation.
In addition, the magnetic field does not change the size of the first core significantly \citep{machida04,machida05a,machida05b}.
Thus, a spatial resolution of at least $\lesssim 1-3$\,AU is expected to be necessary to investigate  disk formation because the first adiabatic core has a size of $\sim 1$\,AU at its formation \citep{saigo06}.

A large-scale disk-like structure is confirmed in Figure~\ref{fig:2}.
The disk corresponds to the so-called ``pseudo-disk" \citep{galli93} because it is not supported by  rotation, as shown in Figure~\ref{fig:1}. 
In the figure, no outflow is confirmed in any model at this epoch.
Although a weak outflow appears just after protostar formation in models US1 -- US4, it gradually weakens and disappears during the Class 0 stage (i.e. $M_{\rm ps}<0.5\msun$).
In general, the outflow driven by the disk (or first core) never disappears even during the main accretion phase \citep{hennebelle09,joos12,seifried12,seifried12b,machida12, machida13}.
Thus, the sink treatment for these models seems to be highly problematic for investigating disk formation and outflow driving.

\begin{figure}
\includegraphics[width=150mm]{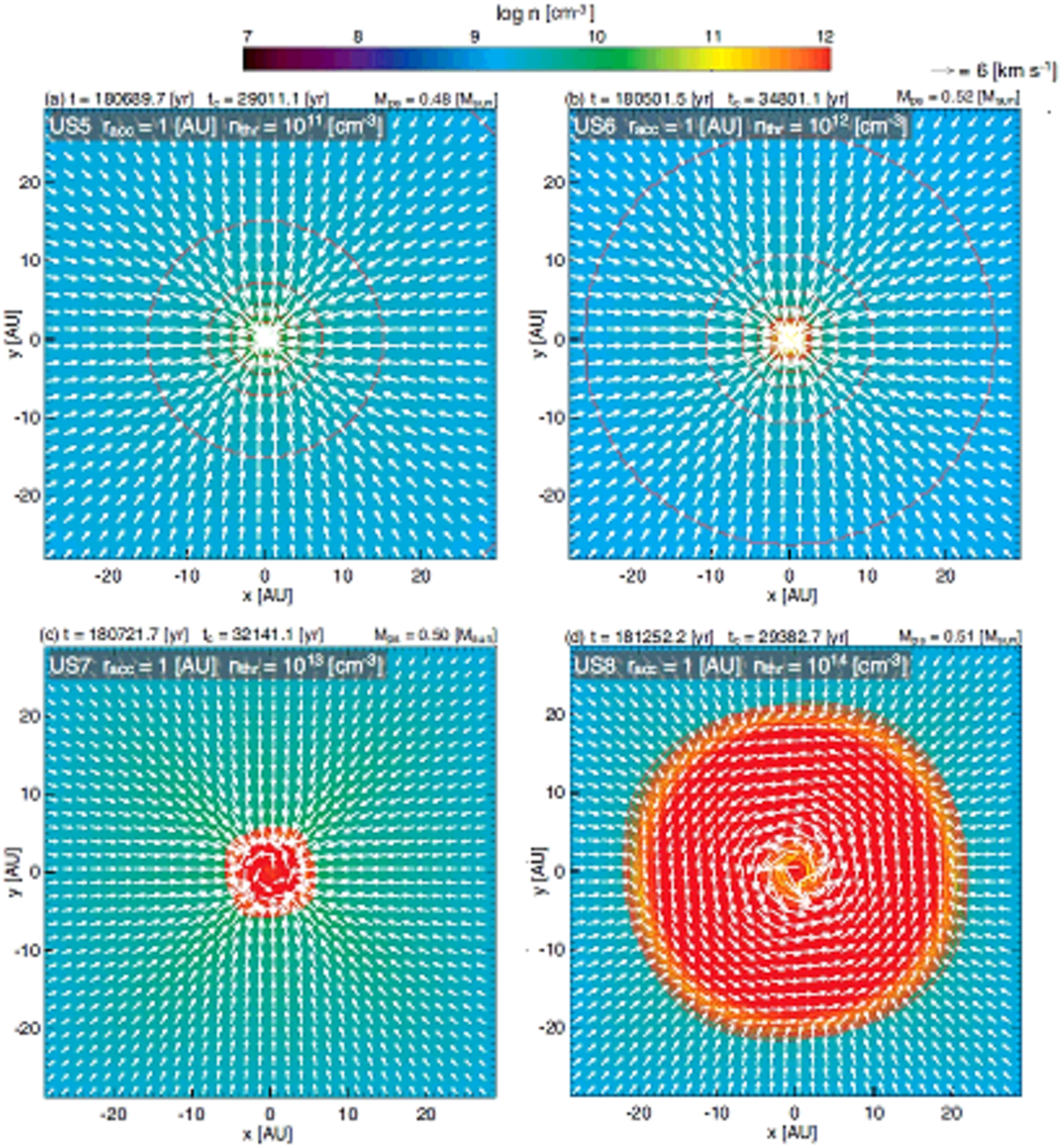}
\caption{
Density (colour and contours) and velocity (arrows) distributions on the equatorial plane for models US5, US6, US7 and US8 when the protostellar mass reaches $M_{\rm ps }\simeq 0.5\msun$.
Elapsed time after the cloud begins to collapse $t$ and that after protostar formation $t_c$ and protostellar mass $M_{\rm ps}$ are given in each panel.
Model name, sink accretion radius $r_{\rm acc}$ and threshold density $n_{\rm thr}$ are also given.
}
\label{fig:3}
\end{figure}

Figure~\ref{fig:3} shows the density and velocity distributions on the equatorial plane when the protostellar mass reaches $M_{\rm ps}\simeq0.5\msun$ for models US5 -- US8, in which the accretion radius is fixed at $r_{\rm acc}=1$\,AU, but the (fixed) threshold density differs among the models.
The disk does not exist at this epoch in models US5 and US6, which have relatively low threshold densities of $n_{\rm thr}=10^{11}\cm$ and $10^{12}\cm$, respectively. 
In model US7, a tiny disk is confirmed in Figure~\ref{fig:3}{\it c}, although it disappears at a later evolutionary stage ($t_c\gtrsim 3.5\times10^4$\,yr).
On the other hand, a clear disk $\sim20$\,AU in size is seen in model US8 (Fig.~\ref{fig:3}{\it d}).
This disk is supported by rotation and does not disappear by the end of the calculation.
In this model, the disk size reaches $\sim30$\,AU at the end of the calculation.

\begin{figure}
\includegraphics[width=150mm]{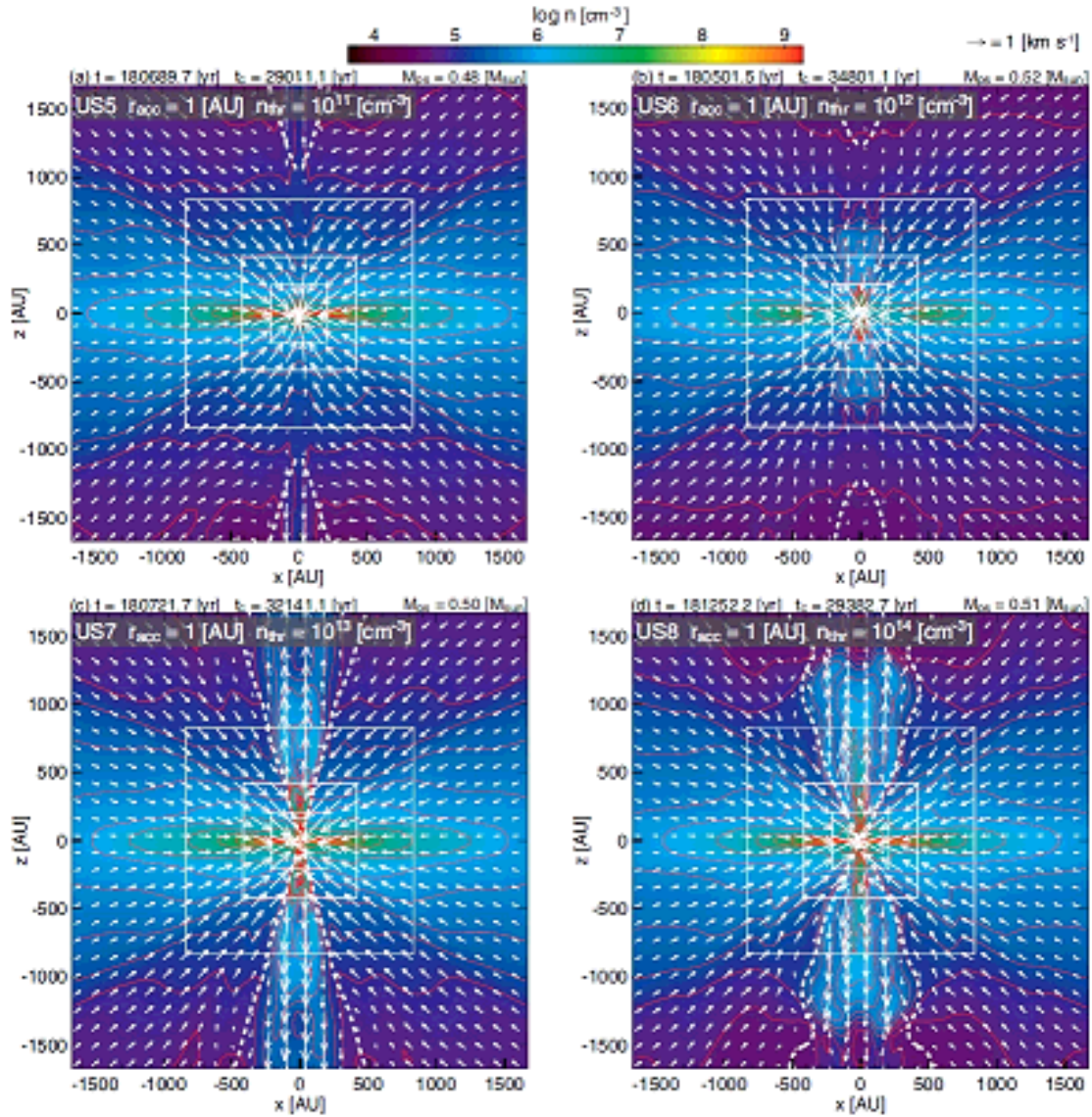}
\caption{
Same as  Fig.~\ref{fig:3}, but on the $y=0$ plane.
Box scale differs from that in Fig.~\ref{fig:3}. 
White dotted line in each panel corresponds to the boundary between infalling gas and outflowing gas, inside of which the gas is outflowing from the center of the cloud.
}
\label{fig:4}
\end{figure}

Figure~\ref{fig:4} shows the density and velocity distributions on the $y=0$ plane for models US5 -- US8.
The boundary between the infalling ($v_r<0$) and outflowing ($v_r>0$) gas is indicated by the white dotted lines, inside of which the gas is outflowing from the centre of the cloud.
In models US1 -- US8, the outflow, which is driven near the protostar, appears just before protostar formation.
However, after protostar formation, the outflow gradually weakens and finally disappears in models US5 and US6 (and US1 -- US4).
The outflow far from the centre of the cloud ($z\gtrsim1000$\,AU) seen in Figures~\ref{fig:4}{\it a} and {\it b} is a remnant of the outflow during the early main accretion stage.
On the other hand, outflow continues to be driven by the centre of the cloud, or by the rotation disk, in models US7 and US8 as shown in Figures~\ref{fig:4}{\it c} and {\it d}, respectively.

\begin{figure}
\includegraphics[width=150mm]{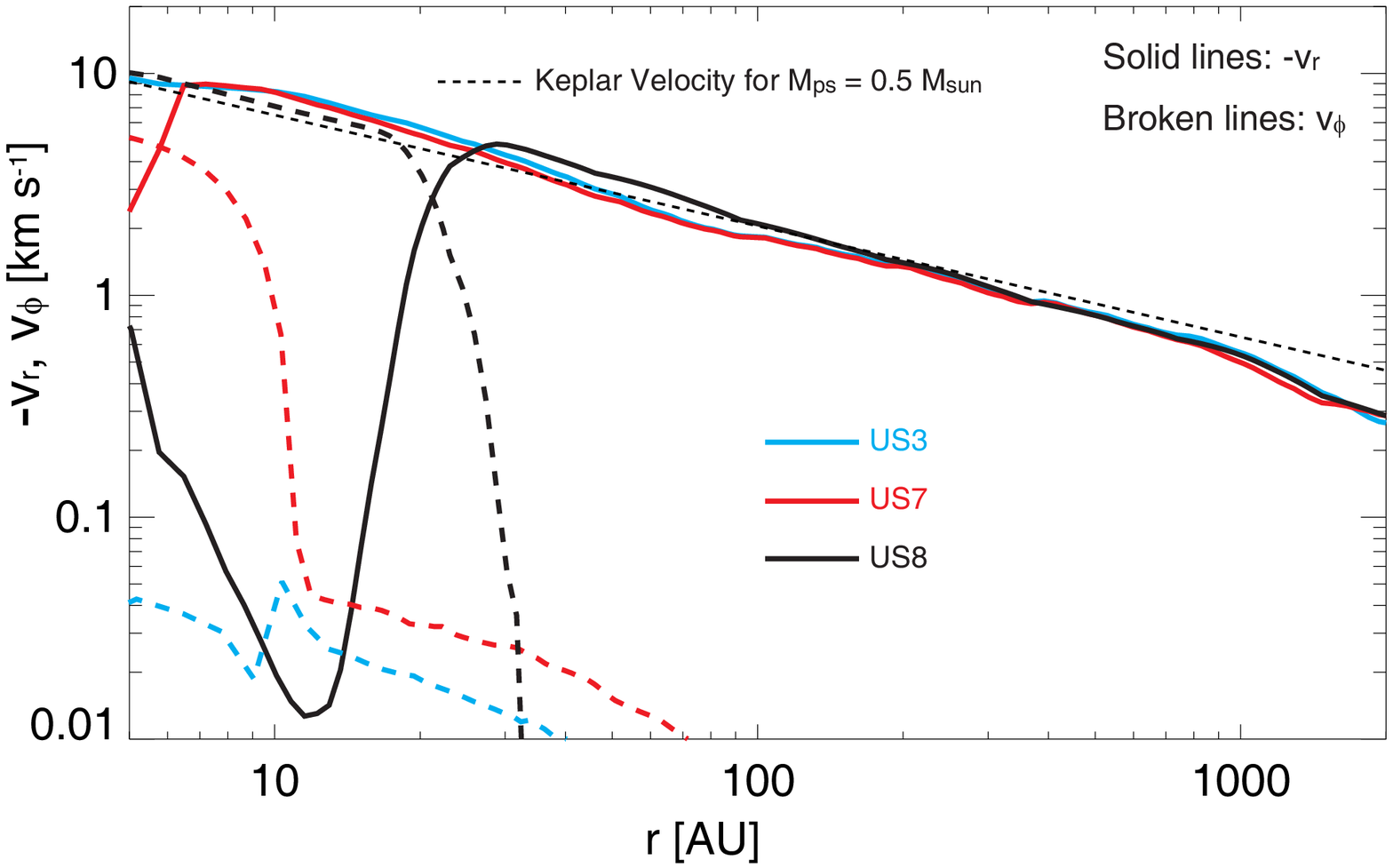}
\caption{
Radial $-v_r$ (solid line) and azimuthal $v_\phi$ (dashed line) velocities at $M_{\rm ps}\simeq0.5\msun$ (the same epoch as in Figs.~\ref{fig:1} and \ref{fig:3}) against the distance from the centre of the cloud for models US3, US7 and US8. 
Keplerian velocity for $M_{\rm ps}=0.5\msun$ is also plotted (dotted line).
}
\label{fig:5}
\end{figure}

Figure~\ref{fig:5} shows the radial and azimuthal velocities, which are azimuthally averaged on the equatorial plane, against the distance from the centre of the cloud for models US3, US7 and US8.
In model US3, the negative radial velocity strongly dominates the azimuthal velocity in the entire region; thus, the gas falls directly  onto the protostar.
On the other hand, in model US8, the negative radial velocity suddenly drops and the azimuthal velocity begins to increase at $r \sim 30$\,AU.
In addition, the azimuthal velocity corresponds well to the Keplerian velocity in the range of $r\lesssim 20$\,AU.
Thus, a Keplerian disk $\sim20$\,AU in size exists at this epoch in model US8. 
In model US7, the azimuthal velocity increases at $r\sim10$\,AU and dominates the radial velocity in the region of $r\lesssim6$\,AU.
However, the azimuthal velocity does not reach the Keplerian velocity.
Thus, the disk is considered to be supported by both rotation and thermal pressure.

\begin{figure}
\includegraphics[width=150mm]{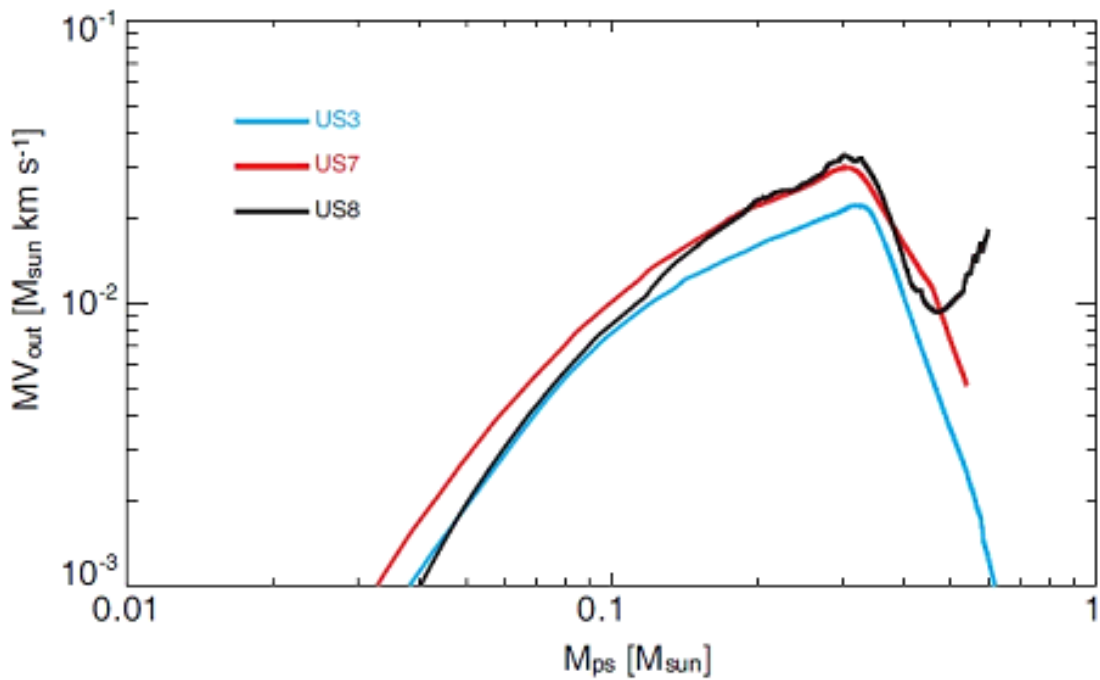}
\caption{
Outflow momentum against the protostellar mass for models US3, US7 and US8.
}
\label{fig:6}
\end{figure}

Figure~\ref{fig:6} plots the outflow momentum against the protostellar mass for models US3, US7 and US8.
In these models, the outflow momentum begins to decrease at $M_{\rm ps}\sim0.3\msun$ and has a peak value of $MV_{\rm out}\sim 0.03 \msun$\,km\,s$^{-1}$.
The peak value of the outflow momentum is considerably smaller than the observations \citep[e.g.][]{curtis10}.
After the peak, the outflow momentum continues to decrease in models US3 and US7, whereas it begins to increase in US8.
Therefore, at the end of the calculation, model US3 has $MV_{\rm out}<10^{-3}\msun$\,km\,s$^{-1}$, whereas model US8 has $MV_{\rm out}\simeq 0.02 \msun$km\,s$^{-1}$.
As seen in Figure~\ref{fig:3}{\it d}, only model US8 shows a Keplerian disk that can continue to drive the protostellar outflow during the gas accretion phase.
Thus, Figures~\ref{fig:4} and \ref{fig:6} indicate that the outflow momentum depends strongly on the sink properties.

\subsubsection{Comparison with Previous Study and Brief Summary}
The cloud parameters for models US1 -- US8 are completely identical to those of the fiducial cloud model of \citet{li11}.
Note that we imposed a computational boundary 16 times that of the initial cloud radius to prevent reflection of Alfv$\acute{e}$n waves, whereas \citet{li11} imposed the computation boundary just outside the initial cloud. 
Also note that both ambipolar diffusion and Ohmic dissipation are included in \citet{li11}, whereas our calculation considers only  Ohmic dissipation.
Although there are slight differences, our result is very similar to \citet{li11}.

Among models US1 -- US8, which have the same initial conditions, we changed only the sink properties of accretion radius $r_{\rm acc}$ and threshold density $n_{\rm thr}$.
The sink condition for model US3 corresponds to that adopted in \citet{li11}, in which a sink accretion radius of $r_{\rm acc}=6.7$\,AU and a variable threshold density (see, \S\ref{sec:sink}) are imposed. 
In model US3 (and US4), a rotation-supported disk never forms during the main accretion phase.
In addition, only a weak outflow appears just after protostar formation, although it gradually weakens and disappears.
These features agree well with \citet{li11}.

On the other hand, a (tiny) disk forms just after protostar formation in models US1, US2 and US5--US8.
This indicates that a sink radius smaller than $r_{\rm acc}\lesssim 3$\,AU is necessary to resolve the disk (or the first core) in the {\it early} main accretion phase.
However, the disk disappears during the main accretion phase for these models except for model US8.
A rotation-supported disk persists until the end of the calculation for model US8.
In addition, during the main accretion phase, the outflow continues to be driven in model US8, whereas it weakens and/or disappears in the other models. 
Thus, it is expected that we need at least an accretion radius of  $r_{\rm acc}\lesssim 1$\,AU and a threshold density of $n_{\rm thr}\gtrsim 10^{13}-10^{14}\cm$ to investigate the long-term evolution of the disk and outflow during the main accretion phase.
The threshold density seems to be more important than the accretion radius for resolving the rotation supported disk, as shown in Figures~\ref{fig:3} and \ref{fig:4}, in which the accretion radius is fixed but the threshold density varies. 
In addition, the calculation results may also depend  on the spatial resolution because different cell widths $h$ are adopted to resolve the accretion radius: the spatial resolution of $h=0.4$\,AU is adopted for models US1 -- US6, whereas $h=0.1$\,AU for model US8 (Table~\ref{table:1}).
Moreover, the calculations do not converge to the same results, as shown in Figures~\ref{fig:1} -- \ref{fig:6}, which indicates that we need  to impose a higher threshold density ($n_{\rm thr}>10^{14}\cm$) and smaller accretion radius ($r_{\rm acc}<1$\,AU) or a greater spatial resolution $h<0.1$\,AU to investigate disk formation {\it in a uniform sphere}.

\subsection{Disk Formation in a Bonnor-Ebert Sphere}
\label{sec:be}
\subsubsection{Initial Cloud Properties and Numerical Settings}
For models BE1 -- BE6 and BEH, the initial cloud has a critical Bonnor--Ebert  density profile \citep{ebert55,bonnor56}. 
Note that the numerical settings and cloud evolution for model BE are described in \S\ref{sec:condition}.
This density profile is characterized by two parameters: the central density $n_{\rm c,0}$ and isothermal temperature $T$.
For Bonnor--Ebert models, we chose $n_{\rm c,0}=10^5\cm$ and $T=10$\,K, which are identical to those of the uniform cloud model (US1 -- US8, see \S\ref{sec:uniform}).
With these parameters, the initial cloud has a radius of $r_{\rm c} = 1.5\times10^4$\,AU.
Because the critical Bonnor--Ebert sphere is in an equilibrium state, we increased the density inside $r<r_{\rm c}$ by a factor of $f$ to promote cloud contraction.
The parameter $f$ is related to the cloud thermal stability $\alpha_0$ as $\alpha_0 = 0.84\,f^{-1}$ \citep{matsu03}.
To investigate the effect of the initial cloud stability on circumstellar disk formation, we chose $f=1.4$ ($\alpha=0.6$; BE1), 1.7 ($\alpha=0.5$; BE2, BE4, BE5 and BE6) and 2.1 ($\alpha=0.4$; BE3). 
Note that the dependence of the magnetic field strength and rotation rate on disk formation for the Bonnor--Ebert cloud was already investigated in \citet{machida11b}. 
Depending on parameter $f$, each model has a mass of  $M_{\rm c} = 2.1\msun$ ($f=1.4$), $2.6\msun$ ($f=1.7$) or $3.2\msun$ ($f=2.1$).
Thus, reflecting the difference in the initial density profile, the clouds with a Bonnor--Ebert density profile are $\sim2-3$ times more massive than those with a uniform density profile, as listed in the third column of Table~\ref{table:1}.
The initial cloud is immersed in a low-density uniform gas with $n_{\rm ISM} = 10^3\cm$, where $n_{\rm ISM}$ is the density of the surrounding gas (or the interstellar medium).
Because we imposed gravity only inside the initial cloud radius,  only the gas at $r<r_{\rm c}$ collapses toward the centre of the cloud  during the calculation.
These conditions are the same as those for the uniform density models.

A uniform magnetic field is imposed in the entire computational domain for models BE1, BE2, BE3, BE4 and BE6, whereas a non-uniform magnetic field is adopted for model BE5.
For model BE5, the magnetic field strength on the equatorial plane is proportional to the square root of the gas density as $B_0 \propto \rho^{1/2}$, in which the plasma beta becomes constant on the equatorial plane.
For models BE1 -- BE5, the magnetic field strength is adjusted to yield the same mass-to-flux ratio of $\mu = 3$, which is almost identical to that in the uniform density models (US1 -- US8).
Thus, the magnetic field strength $B_0$ differs among the models, as listed in the fifth column of Table~\ref{table:1}.
Note that the initial magnetic field strength $B_0$ for model BE5 corresponds to that at the centre of the initial cloud. 
In addition, a relatively strong magnetic field with $\mu=1.7$ is adopted  for model BE6.
For each model, the angular velocity is adjusted to become $\beta_0=0.02$; thus, each cloud has a somewhat different initial angular velocity $\Omega_{\rm 0}$, as listed in the sixth column of Table~\ref{table:1}.
Although it is difficult to construct a Bonnor--Ebert cloud having  the same parameters as a uniform cloud, the mass-to-flux ratio and rotational energy normalized by the gravitational energy for models BE1 -- BE6 are almost the same as those for models US1 -- US8.

The initial cloud with a critical Bonnor-Ebert density profile  is immersed in the $l=5$ grid level, which has a box size of $L(5) = 3.0\times10^4$\,AU.
Thus, the $l=1$ grid has a box size of $L(1)=4.8\times10^5$\,AU.
The maximum grid level is restricted to $l=15$ and has a box size of $L(15)=29$\,AU and a cell width of $h(15)=0.5$\,AU.
These models are identical to those in \citet{machida11b}, although the model parameters and grid size differ somewhat.  
For the sink, only the accretion radius $r_{\rm acc}$ (=3\,AU) is imposed for models BE1, BE2, BE3, BE5 and BE6, in which the threshold density is changed in each timestep and is determined by the density in the cell just outside the sink cell (i.e. a variable threshold density), as described in \S\ref{sec:sink}.
This sink treatment mimics that adopted in \citet{li11}.
Note that an accretion radius of $r_{\rm acc}=6.7$\,AU is adopted in \citet{li11}.
On the other hand, both an accretion radius of $r_{\rm acc} = $3\,AU and a fixed threshold density of $n_{\rm thr} = 10^{12}\cm$ are imposed in model BE4, which corresponds to the sink treatment in \citet{machida11b} and \citet{seifried12}.

\subsubsection{Calculation Results}
\begin{figure}
\includegraphics[width=140mm]{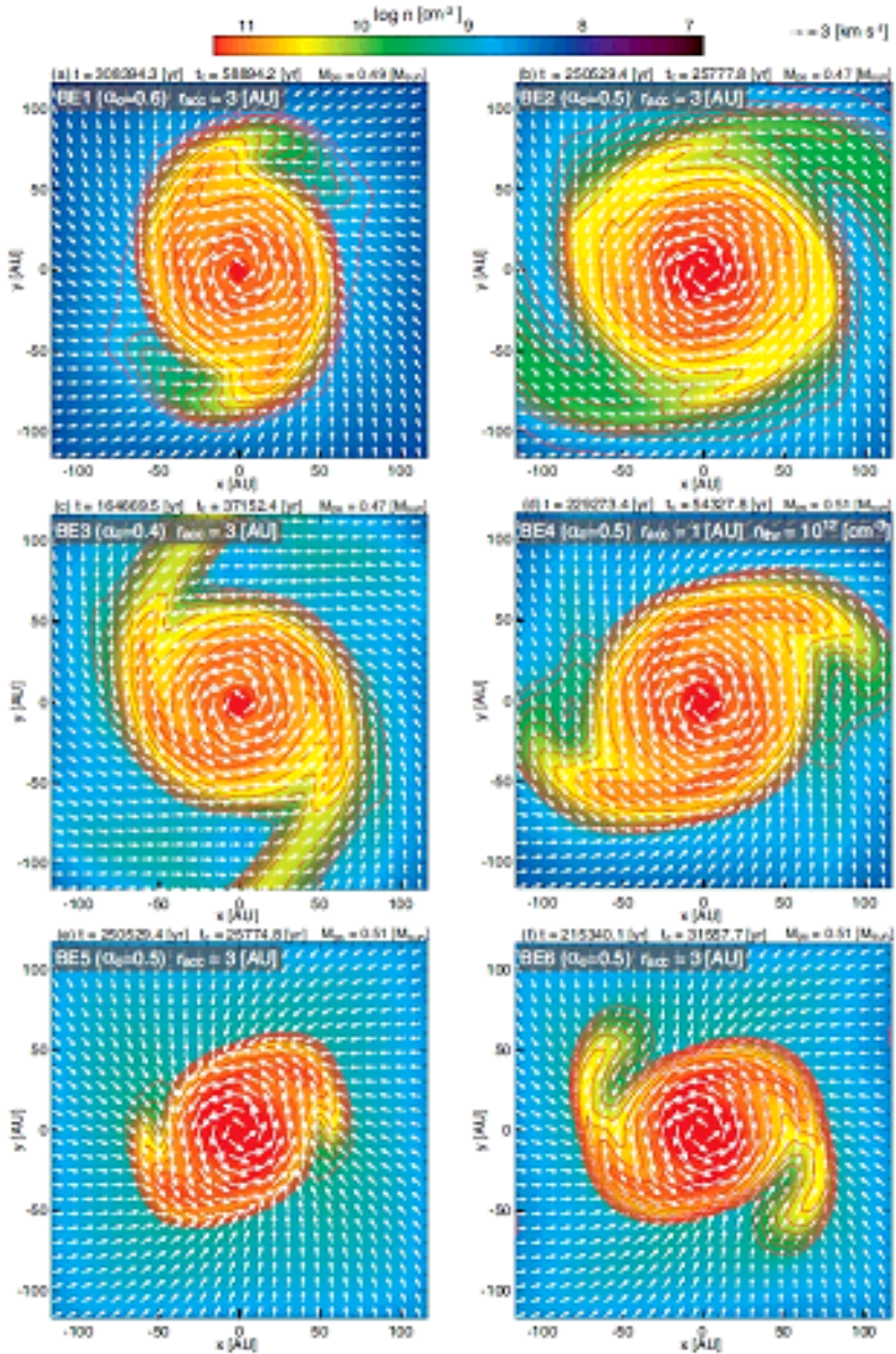}
\caption{
Density (colour and contours) and velocity (arrows) distributions on the equatorial plane for models BE1--BE6 when the protostellar mass reaches $M_{\rm ps }$ approximately $0.5\msun$.
Elapsed time after the cloud begins to collapse $t$ and that after protostar formation $t_{\rm c}$ and protostellar mass $M_{\rm ps}$ are given in the upper part of each panel.
Model name, parameter $\alpha_0$ and accretion radius $r_{\rm acc}$ are also given. 
Threshold density $n_{\rm thr}$ is described in panel (d), in which both an accretion radius and a threshold density are imposed for the sink.
}
\label{fig:7}
\end{figure}

Figure~\ref{fig:7} shows the density and velocity distributions on the equatorial plane when the protostellar mass reaches $M_{\rm ps} \sim0.5 \msun$ for models BE1 -- BE6.
The figure indicates that for each model, a sufficiently large disk of $\sim50-100$\,AU in size exists in the early phase of star formation. 
At this epoch, because the mass of the infalling envelope ($M_{\rm env} > 1\msun$) is much larger than that of the protostar ($M_{\rm ps}\sim0.5\msun$), the protostar is classified as being in the Class 0 stage \citep{andre93,andre94}.
In the figure, a spiral structure is confirmed in all the models.
This structure is considered to be developed by gravitational instability \citep{toomre64} because the disk mass is comparable to the protostellar mass during the (early) Class 0 stage \citep{inutsuka10,inutsuka12}.
A comparison of Figure~\ref{fig:7} with Figures~\ref{fig:1} and \ref{fig:3} indicates a significant difference in disk formation between the Bonnor-Ebert and uniform density cloud models.
A rotating disk appears for the sink with $r_{\rm acc}=3$\,AU and without a fixed threshold density in the Bonnor-Ebert cloud models, whereas no disk appears for the same sink treatment in the uniform cloud models (Fig.~\ref{fig:1}).
The disk continues to grow until the end of the calculation, and its size reaches $\gtrsim 100$\,AU  for the Bonnor-Ebert cloud models.

Although models BE1 -- BE5 have the same mass-to-flux ratio of $\mu = 3$, model BE5 has an initially non-uniform magnetic field.
In addition, model BE6 has the strongest initial magnetic field with $\mu=1.7$.
The disk size in models BE5 and BE6 is somewhat smaller than those in  other models BE1 -- BE4 when the protostar has a mass of  $M_{\rm ps}\simeq0.5\msun$.
On the other hand, the disks seen in Figure~\ref{fig:7} are much larger than those in the uniform cloud models (Fig.~\ref{fig:3}{\it d}), although almost the same initial mass-to-flux ratio is adopted in  both models.
This indicates that the initial cloud configuration or initial density profile strongly affects the disk formation and evolution, whereas the initial distribution and strength of the magnetic field do not significantly affect them.

\begin{figure}
\includegraphics[width=140mm]{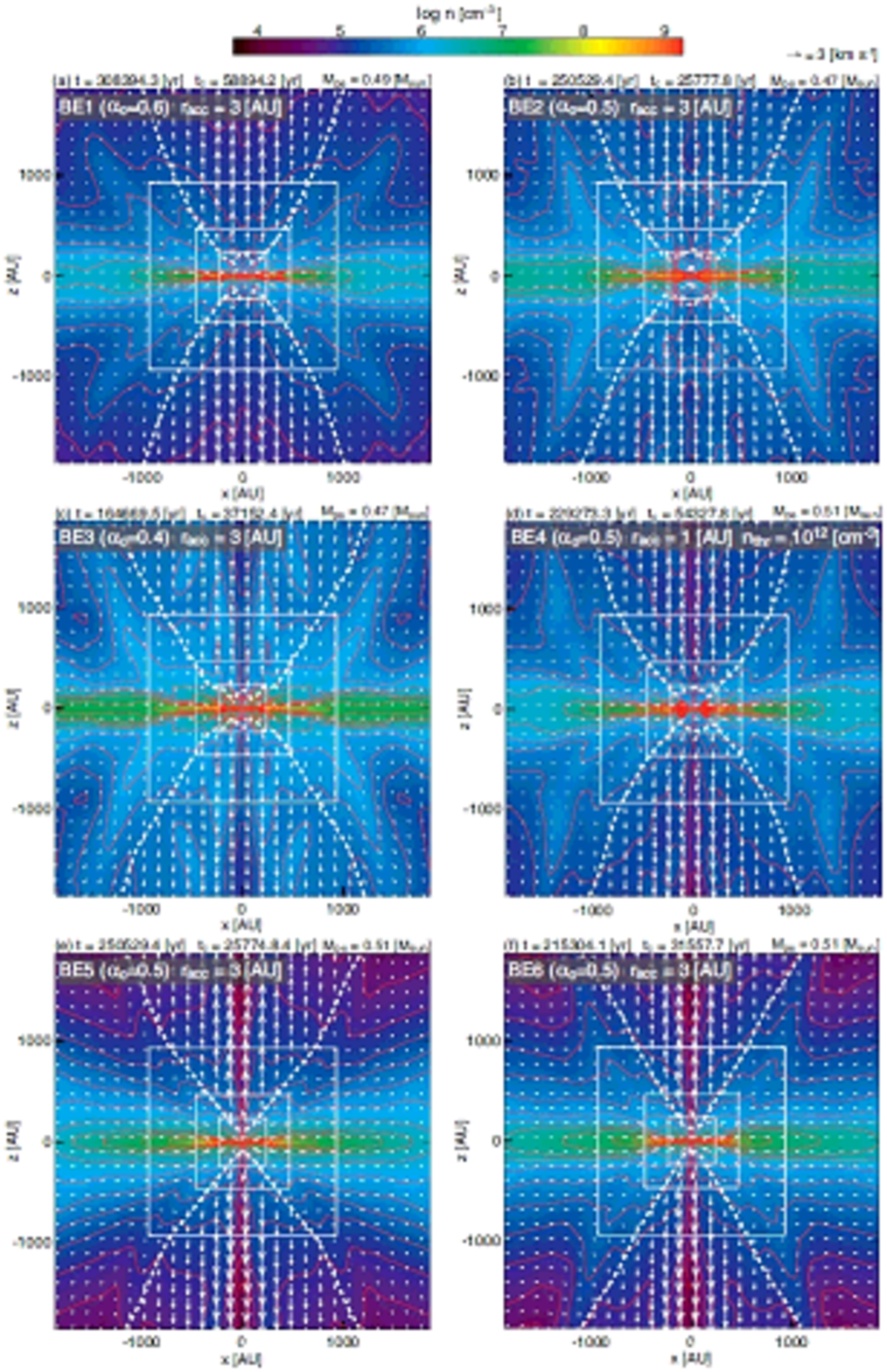}
\caption{
Same as Fig.~\ref{fig:7}, but on the $y=0$ plane.
Box scale differs from that in Fig.~\ref{fig:7}.
White dotted line in each panel corresponds to the boundary between infalling and outflowing gas, inside of which the gas is outflowing from the centre of the cloud with a velocity exceeding sound speed (i.e. $v_r > c_s$).
}
\label{fig:8}
\end{figure}

Figure~\ref{fig:8} shows the density and velocity distributions on the $y=0$ plane for models BE1 -- BE6 at the same epochs shown in Figure~\ref{fig:7}.
The white dotted line in the figure is the boundary between outflowing and infalling gas. 
Strong mass ejection is confirmed in all the models; gas is outflowing near the disk with a wide opening angle. 
In addition, the outflow does not weaken by the end of the calculation. 
Because a weak outflow appears only during the early main accretion phase in the uniform cloud models, the large-scale cloud evolution also differs considerably between the Bonnor-Ebert and uniform cloud models.

\begin{figure}
\includegraphics[width=150mm]{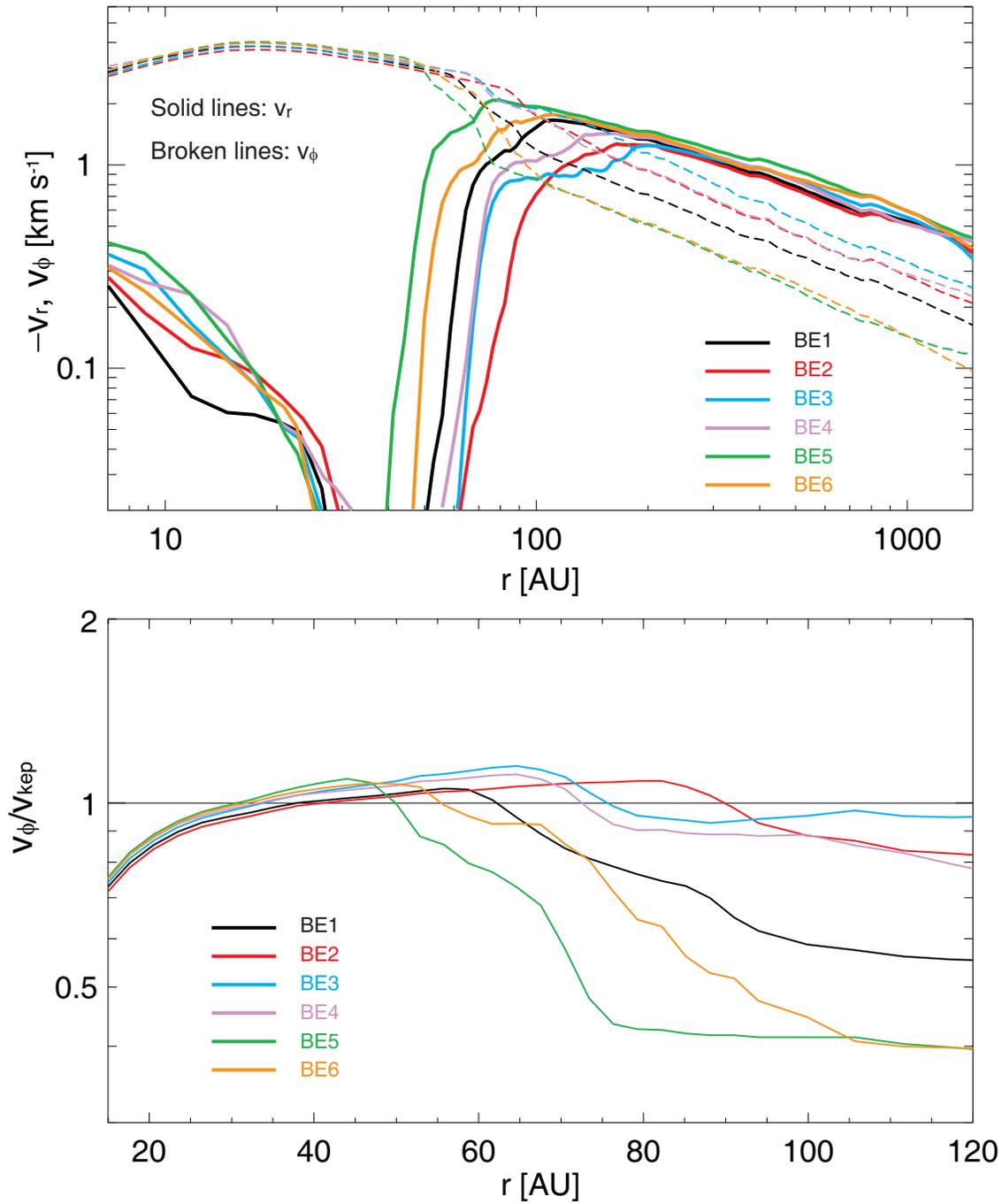}
\caption{
Top: negative radial $-v_r$ (thick solid lines) and azimuthal $v_{\phi}$ (thin broken lines) velocities against the distance from the protostar for models BE1 -- BE6.
Bottom: azimuthal velocity normalized by the Keplerian velocity $v_{\rm kep}$ at 15\,AU $<r<$ 120\,AU; thin horizontal line corresponds to $v_{\phi}=v_{\rm kep}$.
}
\label{fig:9}
\end{figure}

Figure~\ref{fig:9} top panel shows the radial ($-v_r$) and azimuthal ($v_{\phi}$) velocities for all the Bonnor--Ebert models at the same epochs as in Figures~\ref{fig:7} and \ref{fig:8}.
In each model, the radial velocity suddenly drops to  $\vert v_r \vert < 0.1$\,km\,s$^{-1}$ at $\sim50-100$\,AU, which corresponds to the surface of the disk.
On the other hand, the azimuthal velocity continues to increase with decreasing radius and dominates the radial velocity after a sudden increase at $\sim50-100$\,AU. 
All the models have almost the same azimuthal velocity of $v_{\phi}=3-4$\,km\,s$^{-1}$ in the region of $\sim10-100$\,AU. 
This is because the azimuthal velocity in this range is almost identical to the Keplerian velocity, and each model has almost the same protostellar mass of $M_{\rm ps}\simeq0.5\msun$ at this epoch.

The azimuthal velocities differ in the outer region ($r\gg100$\,AU), which corresponds to the infalling envelope.
In this region, the azimuthal velocities in models BE5 and BE6  are two or three times lower than that that in  model BE3.
The magnetic field strengths in models BE5 and BE6 are stronger than that in model BE3, as listed in Table~\ref{table:1}.
Thus, this difference is expected to be caused by magnetic braking in the collapsing cloud.
In addition, in the same region, we can confirm a slight difference in azimuthal velocity among models BE1 -- BE3, where model BE3 has the highest azimuthal velocity.
The initial cloud in model BE3 is thermally the most unstable against gravity among the models.
Thus, it is expected that a rapid cloud collapse can minimize the effect of magnetic braking.

The azimuthal velocity normalized by the Keplerian velocity for each model in the range of $15$\,AU $< r < $ 120\,AU is plotted in Figure~\ref{fig:9} (bottom panel).
The panel indicates that the disk azimuthal velocity is almost the same as the Keplerian velocity near the protostar.
Thus, the disk is supported mainly by rotation.
The size of the Keplerian disk depends little on the cloud parameters.
The Keplerian disk is $\sim 55$\,AU in size in model 6, which  has the smallest mass-to-flux ratio ($\mu = 1.7$), whereas the disk size is $\sim50$\,AU in model BE5, which has the strongest initial (non-uniform) magnetic field.
In addition, the disk size in model BE3, which has the smallest $\alpha_0$, is $\sim100$\,AU, whereas model BE1, which  has the largest $\alpha_0$, has a disk size of $\sim 65$\,AU.

\begin{figure}
\includegraphics[width=150mm]{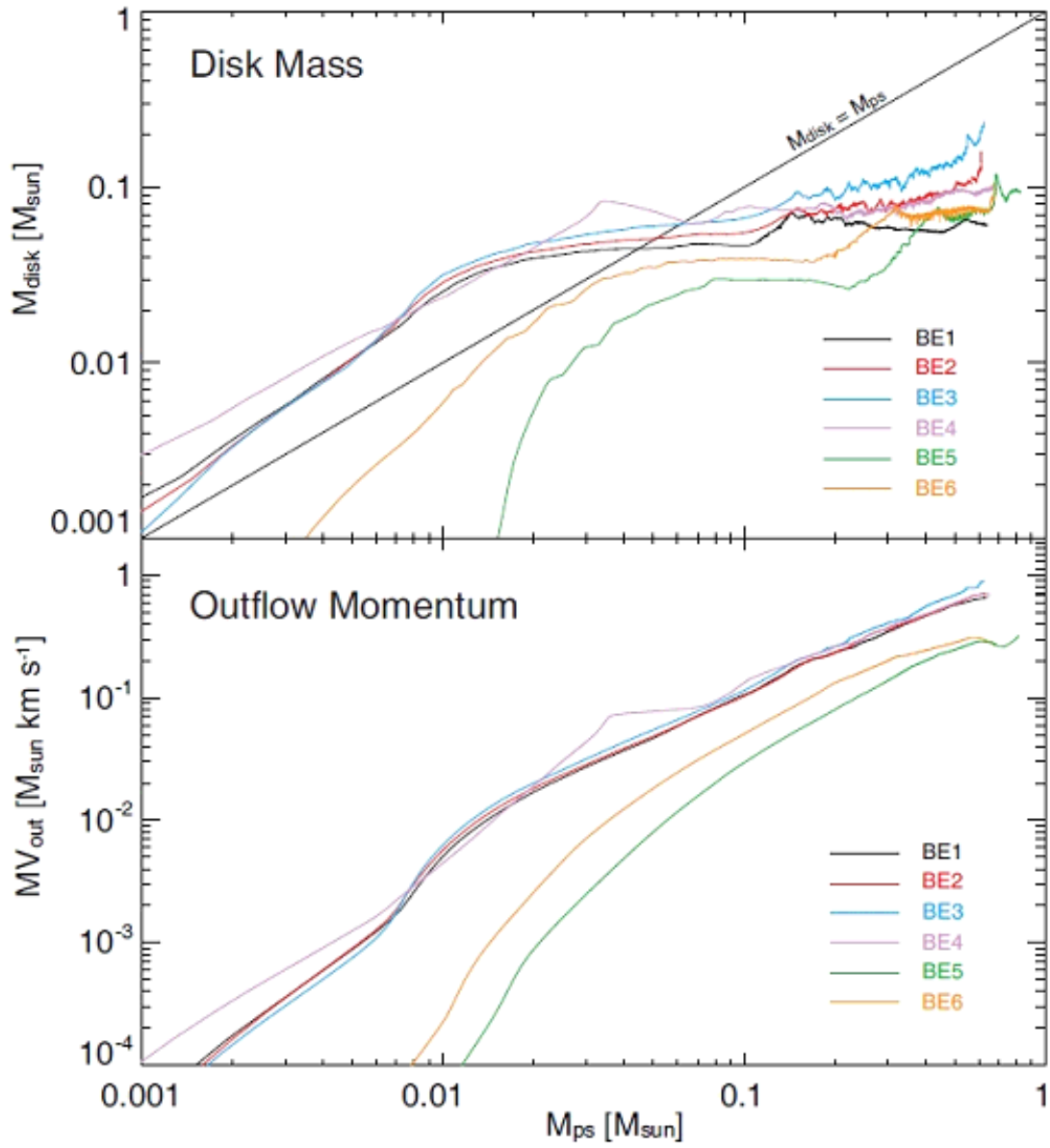}
\caption{
Disk mass (top panel) and outflow momentum (bottom panel)  against the protostellar mass for models BE1 -- BE6.
The relationship $M_{\rm disk}=M_{\rm ps}$ is also plotted (thin solid line in top panel). 
}
\label{fig:10}
\end{figure}

The disk mass is plotted against the protostellar mass in Figure~\ref{fig:10} (top panel). 
The disk mass dominates (models BE1 -- BE4) or is comparable to (models BE5 and BE6)  the protostellar mass in the early phase of star formation. 
At the protostar formation epoch, because the first core becomes the circumstellar disk, the disk mass dominates the protostellar mass, as explained in \citet{inutsuka10}.
In the very early phase of star formation, the disk mass is adjusted by the gravitational instability and it gradually settles into a stable state. 
At the end of the calculation, the disk has  10-30\% of the protostellar mass.
This panel also indicates that models with a smaller $\alpha_0$ or weaker magnetic field have a massive disk.

The outflow momentum is plotted against the protostellar mass in Figure~\ref{fig:10} (bottom panel). 
The outflow momentum continues to increase until the end of the calculation in the Bonnor-Ebert models, whereas it begins to decrease in the  uniform density models.
At the end of the calculation, the outflow for models BE1 -- BE6 has a momentum of $\sim 0.2-1\msun$\,km\,s$^{-1}$.
Observations have shown that Class 0 protostars have outflow momenta of $MV_{\rm out}\sim 0.1-1\msun$\,km\,s$^{-1}$ (\citealt{curtis10}, see also \citealt{bontemps96, wu04, hatchell07}).
Thus, the outflow momentum in models BE1 -- BE6 corresponds well to the observations.
In addition, the panel indicates that the outflow momentum depends little on the initial magnetic field strength.
Models BE5 and BE6, which have a weaker initial magnetic field, have a smaller outflow momentum. 
However, the difference in outflow momentum among the models at $M_{\rm ps}\sim 0.5\msun$ is only about a factor of five.

\subsubsection{Comparison with Previous Study and Brief Summary}
The evolution of clouds with an initial Bonnor-Ebert density profile was investigated in this section.
To the extent possible, we adopted the same cloud parameters for Bonnor-Ebert cloud models as for the uniform sphere models \citep{li11}, in which almost the same initial density (at the centre of the cloud), mass-to-flux ratio and rotation rate (or ratio of rotational to gravitational energy) were used.
Note that the cloud mass and radius for the Bonnor-Ebert models differs from those of the uniform cloud models because we cannot construct an initial cloud with exactly the same parameters in models with different density distributions.
However, we adopted other cloud parameters (or other initial densities) in our previous study \citep{machida11b}, in which typical clouds have almost the same mass and radius as in the uniform cloud models (see Table~1 of \citealt{machida11b}).
Thus, using the results of our present and previous studies, we can compare Bonnor-Ebert clouds with the uniform cloud models.

A rotation supported disk $\sim50-100$\,AU in size appears in the Bonnor-Ebert cloud models, whereas a very tiny disk $\sim20$\,AU in size appears in the uniform cloud model. 
Thus, although both clouds have almost the same parameters, the disk size considerably differs.
In addition, we showed that different initial density distributions affect the protostellar outflow driven by the circumstellar disk.
The difference in the outflow momentum among the Bonnor-Ebert and uniform cloud models is more than  one order of magnitude.
During the main accretion phase, the outflow rapidly weakens in the uniform density model, whereas it continues to be driven in the Bonnor-Ebert models. 
These results indicate that the initial cloud configuration significantly affects the early phase of star formation.

The necessary spatial resolution and sink conditions seem to differ among the models.
For the uniform cloud models, a sink accretion radius of $r_{\rm acc}\lesssim 1$\,AU and threshold density of $n_{\rm thr}\gtrsim10^{13}-10^{14}\cm$ are necessary for disk formation.
However, note that the calculation results did not converge among uniform cloud models with different sink conditions.
On the other hand, a rotation-supported disk forms with $r_{\rm acc}=3$\,AU and a variable threshold density in Bonnor-Ebert models BE1, BE2, BE3, BE5 and BE6.
The disk also forms with $r_{\rm acc}=1$\,AU and $n_{\rm thr}=10^{13}\cm$ in model BE4 and the models in \citet{machida11b}, in which cell widths of $h = 0.5$\,AU and $0.6$\,AU, respectively, are adopted.

Models BE2 and BE4 have the same cloud parameters, but the sink conditions differ.
Figure~\ref{fig:10} indicates that the evolution of the disk mass and outflow momentum converge to almost the same values between the models.
This indicates that the spatial resolutions and sink conditions adopted in this study for the Bonnor-Ebert clouds are sufficient for investigating disk formation and outflow driving.

We also focused on the effect of the initial cloud stability and distribution of the magnetic field on disk formation in the Bonnor-Ebert cloud models. 
A high mass accretion rate onto the disk or protostar is realized in an unstable cloud that has a smaller $\alpha_0$.
As shown in Figure~\ref{fig:9}, the cloud with a high mass accretion rate tends to form a relatively massive and large disk. 
However, the difference is not very large.
Instead, the different initial cloud configurations cause large differences in disk formation.

\subsection{Disk Formation in Spherical Clouds with a Steeper Density Profile}
\subsubsection{Initial Cloud Properties and Numerical Settings}
The density distribution of the initial cloud significantly affects circumstellar disk formation, as shown in \S \ref{sec:uniform} and \ref{sec:be}.
To further investigate the effects of the initial configuration, we constructed initial clouds (RJ1 -- RJ4) with the density distribution used in \citet{hennebelle09} and \citet{joos12}.
The density profile of the initial cloud for models RJ1 -- RJ 4 is described as
\begin{equation}
\rho = \dfrac{\rho_0}{1+(r/r_0)^2},
\end{equation}
where $ \rho_0 = 3 \times 10^{-17}$\,g\,cm$^{-3}$ ($n_{\rm c} = 8 \times 10^6\cm$) and $r_0=970$\,AU are adopted.
The mass and radius of the initial cloud are $M_{\rm c} = 1\msun$ and $r_{\rm c}=3\times10^3$\,AU, respectively. 
In addition, for these models we adopted a uniform magnetic field and rigid rotation.
The mass-to-flux ratio for these models is $\mu=3$ and the ratio of the rotational to gravitational energy is $\beta=0.03$. 
The other parameters are listed in Table~\ref{table:1}. 
The cloud properties are almost identical to those in \citet{hennebelle09} and \citet{joos12}.
In addition, the cloud parameters are almost identical to those of the uniform density  and Bonnor-Ebert cloud models.

For models RJ1 -- RJ4, the initial cloud is set to the $l=5$ grid level, which has a box size of $L(5)=6\times10^3$\,AU, and the maximum grid is restricted to $l=13$.
Thus, the first level of the grid has a box size of $L(5)=10^5$\,AU, whereas the maximum grid level  has $L(13)=23$\,AU with a cell width of $0.4$\,AU.
The spatial resolution of the finest grid is comparable to that in \cite{joos12}, in which a maximum spatial resolution of $\sim0.5$\,AU was adopted.

Models RJ1 -- RJ4 have the same cloud parameters but different sink conditions.
\citet{hennebelle09} and \citet{joos12} calculated the cloud evolution without either a sink or magnetic dissipation. 
To mimic this, we did not impose the sink for models RJ1 and RJ3, and did not include Ohmic dissipation for models RJ1 and RJ2.
For the sink models (RJ2 and RJ4), model RJ2 has $r_{\rm acc}=3$\,AU and a variable threshold density, whereas model RJ4 has $r_{\rm acc}=1$\,AU and $n_{\rm thr}=10^{13}\cm$.
The other numerical settings are the same as in the uniform and Bonnor-Ebert cloud models.

\subsubsection{Calculation Results}
\begin{figure}
\includegraphics[width=150mm]{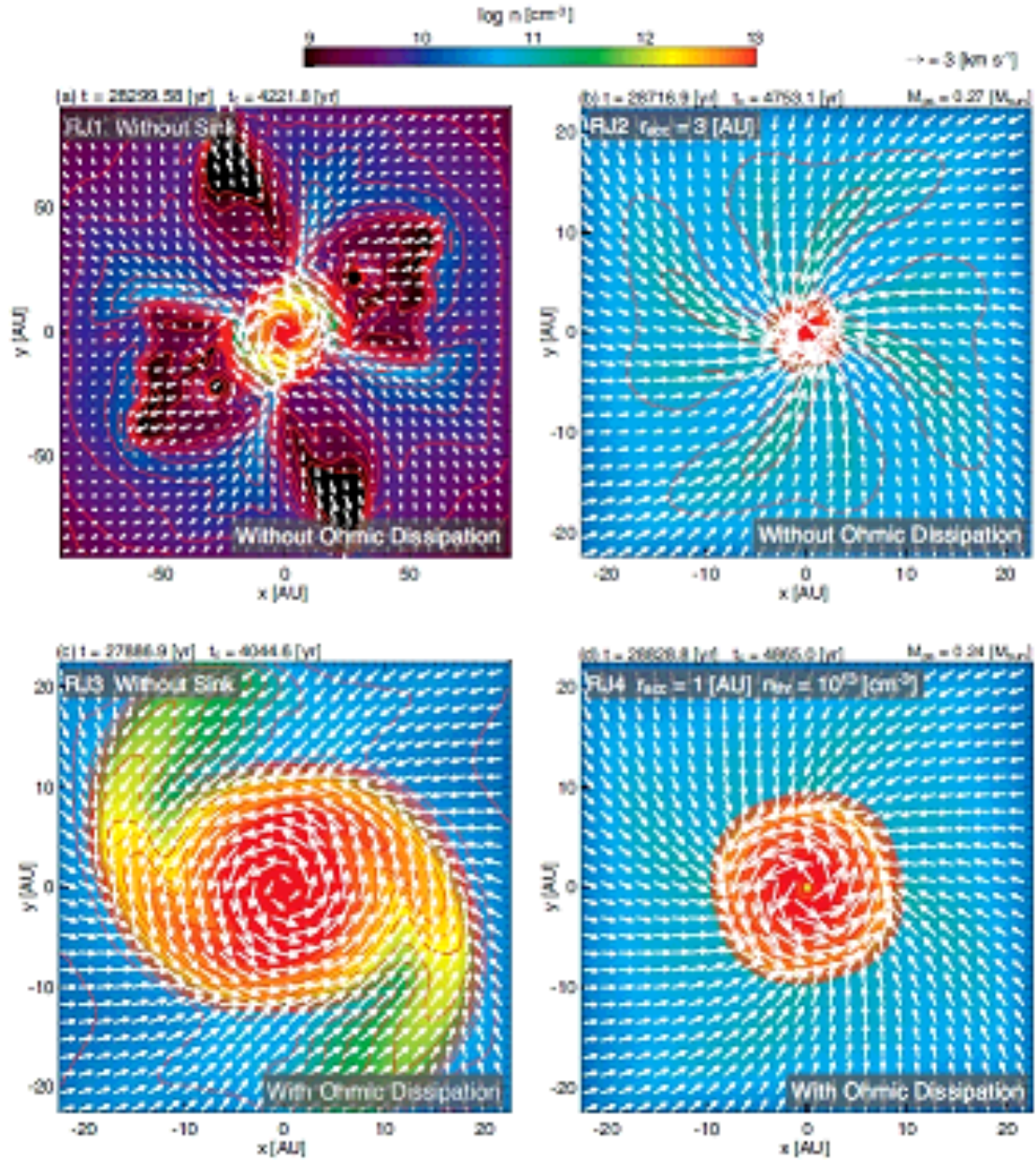}
\caption{
Density (colour and contours) and velocity (arrows) distributions on the equatorial plane for models RJ1 -- RJ4 at $t_{\rm c}$ approximately $4000-5000$\,yr. 
Box size of panel {\it a} differs from those of panels {\it b} -- {\it d}.
Elapsed time after the cloud begins to collapse $t$ and that after protostar formation $t_{\rm c}$ and protostellar mass $M_{\rm ps}$ (models RJ2 and RJ4) are given  in the upper part of each panel.
Model name, sink condition and whether or not Ohmic dissipation is included are given in each panel.}
\label{fig:11}
\end{figure}

Figure~\ref{fig:11} shows the density and velocity distributions on the equatorial plane during $t_c \sim 4000$ -- 5000\,yr after protostar formation for models RJ1 -- RJ4; the box size of each panel differs to emphasize the structure around the protostar or disk. 
Because we cannot define the protostellar mass for models without a sink (RJ1 and RJ3), we did not describe it in panels {\it a} and {\it c}.
In addition, for these models, we defined the protostar formation epoch $t_c=0$ as that at which the maximum (or central) density reaches $n=10^{13}\cm$.
The figure indicates that the disk size strongly depends on both the sink condition and magnetic dissipation process.

A disk like structure $\lesssim 10$\,AU in size appears in model RJ1, which has the  same numerical conditions (without either a sink or Ohmic dissipation) as in \citet{hennebelle09} and \citet{joos12}. 
The disk is surrounded by four cavities within which the gas flows out toward the centre of the cloud. 
The same structure is confirmed in Fig.~19 of \citet{joos12}. 
We discuss the cavity and equatorial outflow in \S\ref{sec:resolution}.
We confirmed that for model RJ1, the thermal pressure greatly contributes to support the disk in addition to the rotation, which is consistent with \citet{hennebelle09} and \citet{joos12}.
Although we cannot correctly define the Keplerian velocity because the protostellar mass is not defined without the sink, we estimate the disk mass as $\lesssim 0.01\msun$ according to the prescription in \citet{machida10a}.
The disk mass is also consistent with that for the same model in \citet{hennebelle09} and \citet{joos12}.

Without Ohmic dissipation and with the sink (model RJ2), no rotation-supported disk forms and only  a very small thermally supported disk appears, as shown in Figure~\ref{fig:11}{\it b}.
In both models RJ1 and RJ2, the disk is supported mainly by thermal pressure, not by rotation.
With the sink, the thermal pressure is removed from the centre of the cloud.
Thus, it is natural that the thermally supported disk is smaller in the model with a sink (model RJ2) than in that without a sink (model RJ1).  
Note that without the sink, the spatial resolution can affect the size of the thermally supported disk because the central (or maximum) density depends on the local spatial resolution,  and the thermal pressure is related to the density distribution around the centre of the cloud.
We confirmed that without the sink, the disk size depends on the local spatial resolution.

A rotation-supported disk forms in models RJ3 and RJ4, both of which include Ohmic dissipation.
The sink is not imposed in model RJ3, which has the largest disk among models RJ1 -- RJ4, as shown in Figure~\ref{fig:11}.
Without the sink, an artificially high gas pressure prevents the cloud from further collapse, and the high-density gas region expands.
Because the magnetic field dissipates in the high-density gas region, the magnetic field weakens around the centre of the cloud. 
Then, the efficiency of magnetic braking also weakens and the rotation-supported disk forms and expands in the collapsing cloud. 
In addition, because a massive disk forms in model RJ3, a spiral structure develops owing to gravitational instability, as shown in Figure~\ref{fig:11}{\it c}. 
Model RJ4 has a smaller disk than model RJ3 because a relatively low gas pressure, which is realized with the sink, produces a smaller thermally supported disk (i.e. the first core or the remnant of the first core) that is the origin of the rotation-supported disk.
Thus, we may overestimate the disk size without the sink, or underestimate it with the sink.

\begin{figure}
\includegraphics[width=150mm]{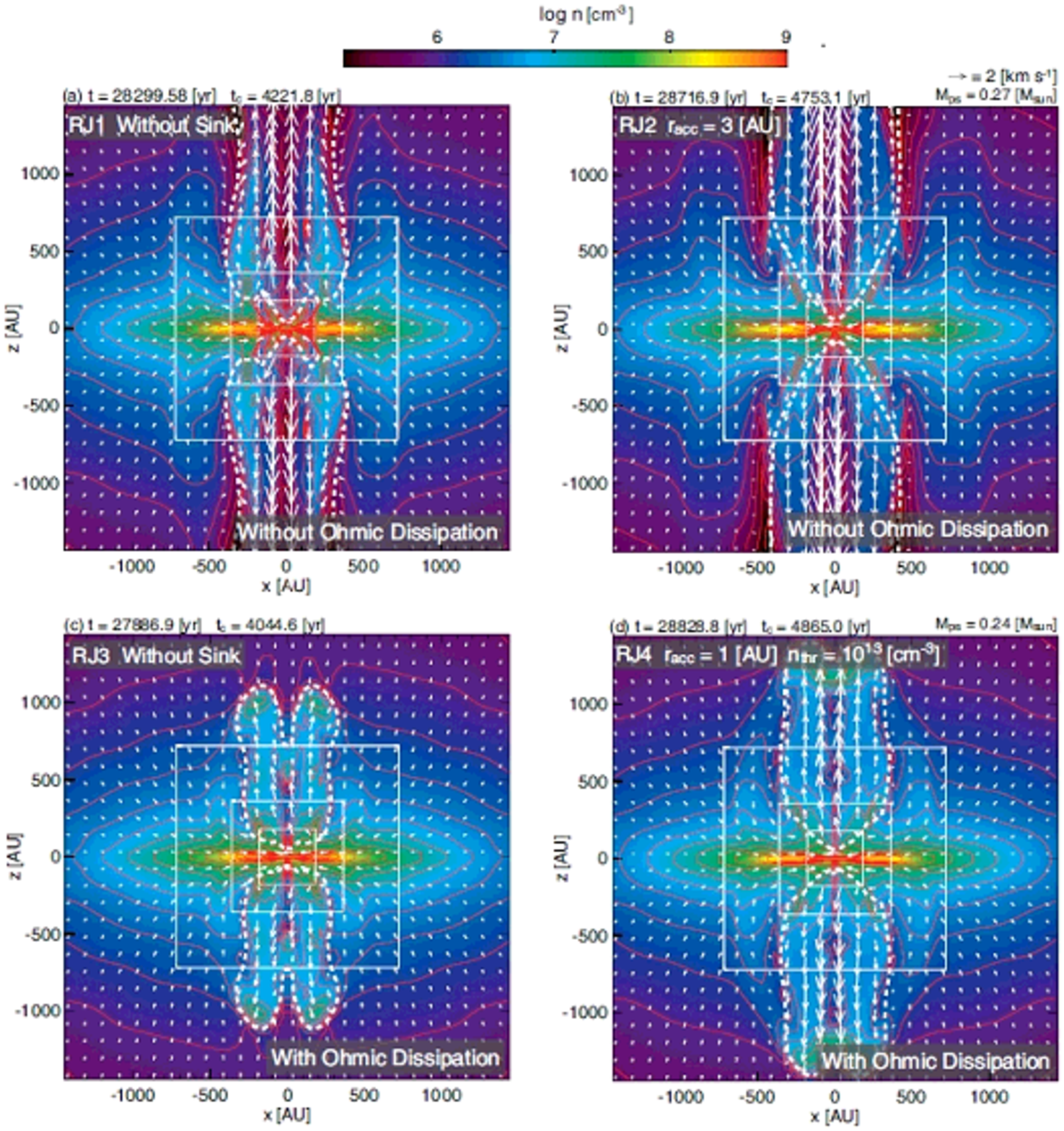}
\caption{
Same as  Fig.~\ref{fig:11}, but on the $y=0$ plane.
Box scale differs from that in Fig.~\ref{fig:11}.
White dotted line in each panel corresponds to the boundary between infalling gas and outflowing gas, inside of which the gas is outflowing from the centre of the cloud with a velocity exceeding sound speed (i.e. $v_r > c_s$).
}
\label{fig:12}
\end{figure}

Figure~\ref{fig:12} shows the density and velocity distributions on the $y=0$ plane at the same epochs as in Figure~\ref{fig:11} for models RJ1 -- RJ4. 
The figure indicates that although the same initial state is imposed among the models, the large-scale structure differs. 
A strong outflow with a relatively wide opening angle appears in model RJ2, whereas a weak outflow with a narrow opening angle appears in model RJ3. 
The models without Ohmic dissipation (RJ1 and RJ2) tend to show a strong outflow, which indicates that the stronger magnetic field around the centre of the cloud drives a stronger outflow.
In addition, comparison of Figure~\ref{fig:8} and \ref{fig:12} indicates that the outflows in the Bonnor-Ebert density models have a wider opening angle than those in model RJ1 -- RJ4.
Thus, different initial density distribution also affects the outflow properties.

\begin{figure}
\includegraphics[width=150mm]{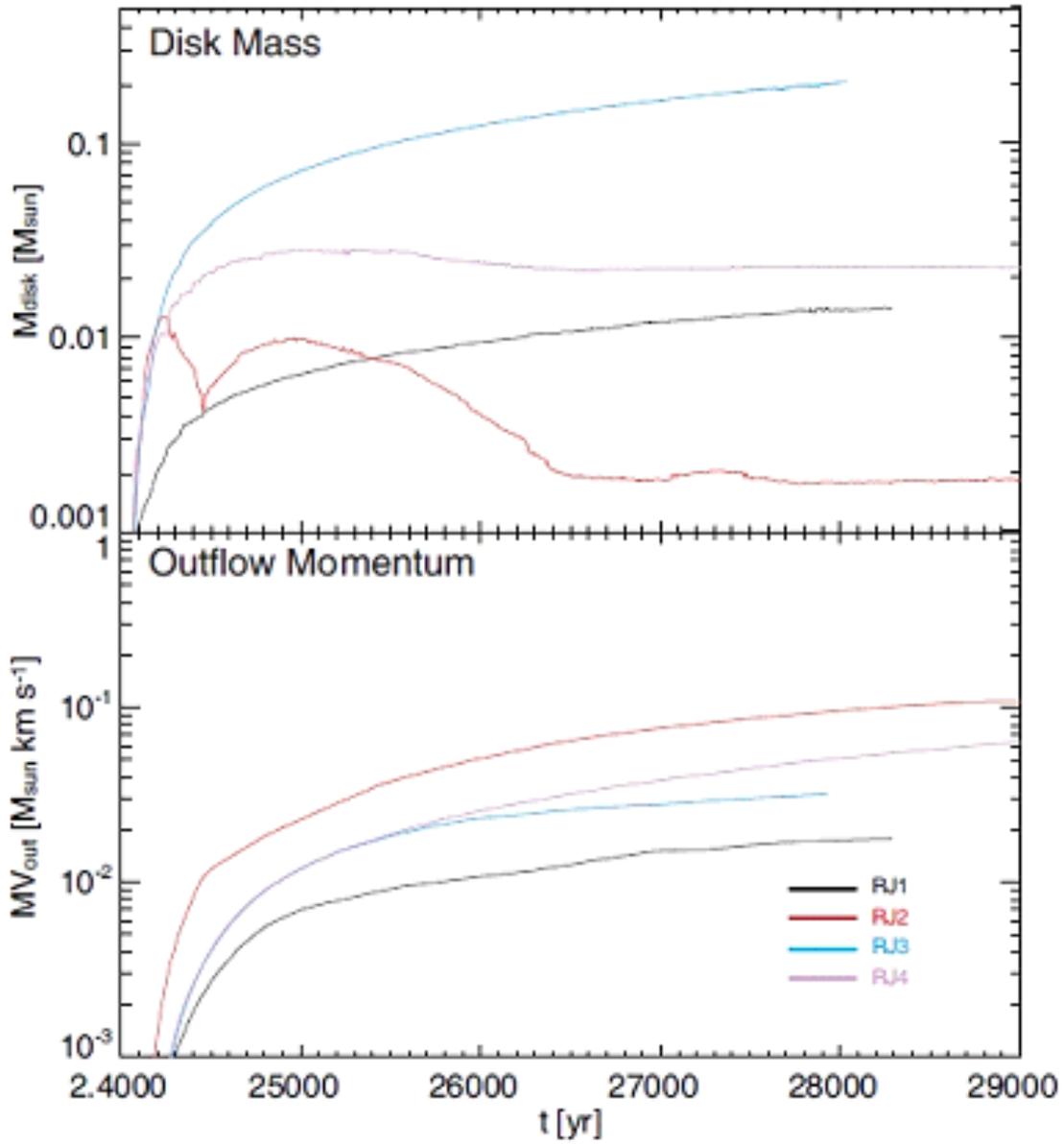}
\caption{
Disk mass (top panel) and outflow momentum for models RJ1 -- RJ4 against the elapsed time after the cloud begins to collapse
}
\label{fig:13}
\end{figure}

The disk mass and outflow momentum are plotted against the time after the cloud begins to collapse in Figure~\ref{fig:13}.
Among the models, the difference in the disk mass is about two orders of magnitude at $t\sim2.8\times 10^4$\,yr, whereas that in the outflow momentum is about one order of magnitude at the same epoch.
Thus, the sink and magnetic dissipation greatly affect the evolution of the disk and outflow.

\subsection{Disk Formation in Massive Clouds}
\subsubsection{Initial Cloud Properties and Numerical Settings}

\citet{seifried12} investigated disk formation in a massive cloud (or disk formation around a massive star).
They adopted the density profile of $\rho(r) \propto r^{-1.5}$ as the initial state.
Note that in their study,  $\rho(r) \propto (1-(r/r_0)^2)$ was used around the centre of the cloud  to avoid an unphysically high density.
As a massive cloud model, we constructed the same initial cloud as in \citet{seifried12}.
The initial density at the centre of the cloud is set to $n_{\rm c,0}=6\times10^6\cm$ with $r_0=0.018$\,pc, and an isothermal temperature of $20$\,K is imposed.
The cloud has a mass of $100\msun$ and a radius of $2.6\times10^4$\,AU (or a diameter of 0.25\,pc) surrounded by low-density gas.
To realize a constant plasma beta on the equatorial plane, the initial magnetic field is set to be proportional to $B_z \propto r^{-0.75}$ \citep{seifried12}.
The initial magnetic field strength at the centre of the core is set to 659$\,\mu$\,G.
The cloud has a mass-to-flux ratio of 5.2.
Initial rigid rotation with an angular velocity of $\Omega_0 = 3.16 \times1 0^{-13} $\,s$^{-1}$ is adopted and the rotation axis is set to be parallel to the magnetic field (or $z$-axis). 
The  cloud model is exactly the same as in \citet{seifried12}, and the cloud parameters are almost identical to model 5.2-4 of that paper.

We constructed four models (models RS1 -- RS4) as massive cloud model.
They have the same cloud parameters but different sink treatments and magnetic dissipation processes, as listed in Table~\ref{table:1}. 
Initially, for all the models, we prepared seven grid levels ($l=1-7$).
The first grid level is $3.1\times10^5$\,AU in size with a cell width of $4.8\times 10^3$\,AU.
Thus, the $l=1$ grid has a box size of about 12 times larger than the initial cloud radius.
The seventh grid level has a box size of $4.8\times10^3$\,AU with a cell width of $76$\,AU.
For model RS1, the maximum grid level is restricted to $l=11$ and has a cell width of 4.7\,AU, which is identical to the finest resolution adopted in \citet{seifried12}.
On the other hand, for models RS2 -- RS4, the maximum grid level is restricted to $l=14$, which has a cell width of $0.6$\,AU.
Thus, the finest resolutions differ by a factor of eight among the models.

The sink is imposed for RS1 -- RS4, but the sink accretion radius and threshold density differ, as listed in Table~\ref{table:1}.
Model RS1 has an accretion radius of $r_{\rm acc}=12.6$\,AU and threshold density of $n_{\rm thr}=3\times10^{11}\cm$.
The sink treatment for model RS1 is the same as in \citet{seifried12}.
Models RS2 and RS4 have $r_{\rm acc}=1$\,AU and $n_{\rm thr}=10^{13}\cm$, whereas model RS3 has $r_{\rm acc}=3$\,AU and $n_{\rm thr}=10^{12}\cm$.
Ohmic dissipation is included for models RS3 and RS4, but not for models RS1 and RS2.
Note that \citet{seifried12} calculated disk formation using the ideal MHD equations.

\subsubsection{Calculation Results}
\begin{figure}
\includegraphics[width=150mm]{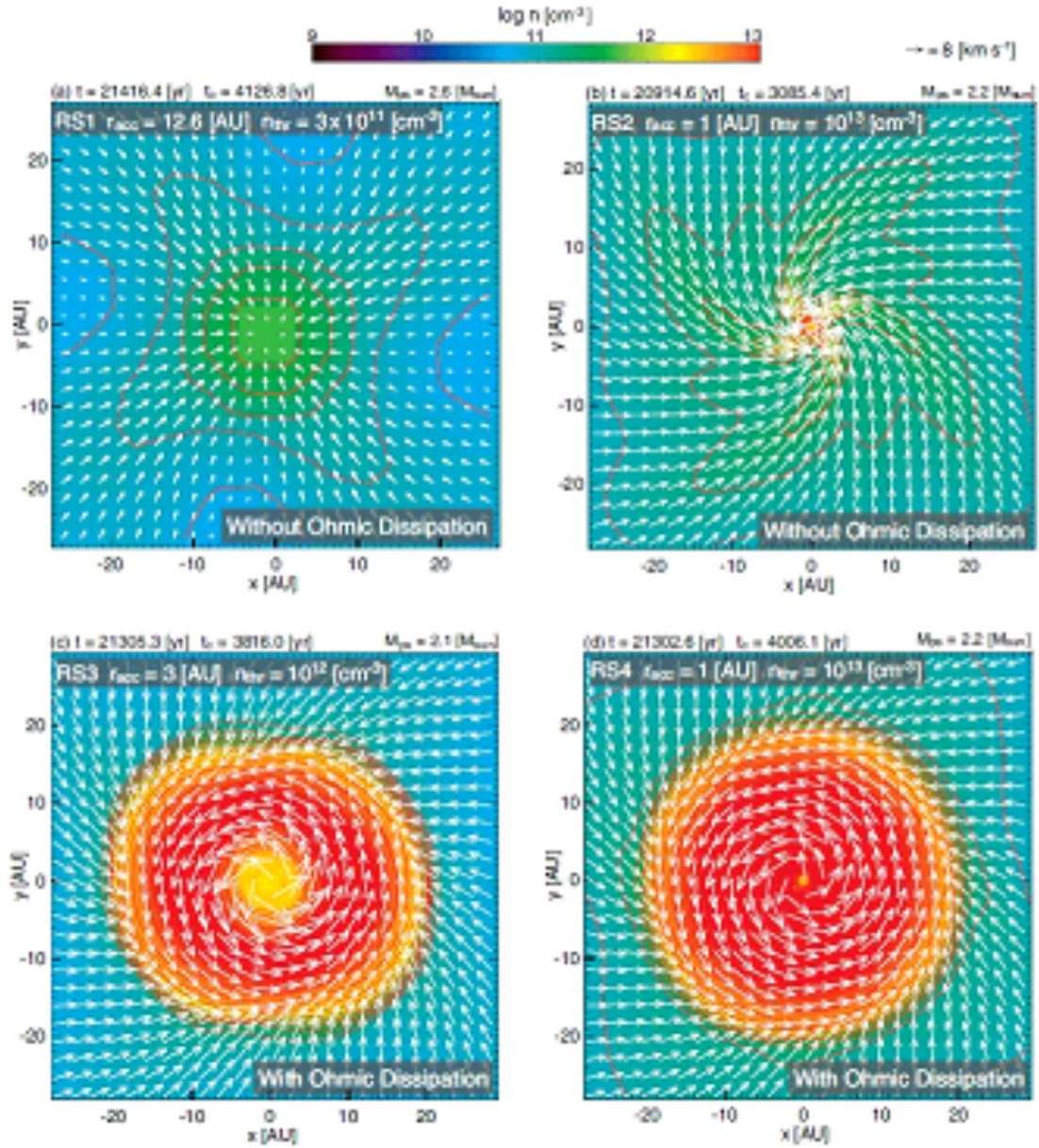}
\caption{
Density (colour and contours) and velocity (arrows) distributions on the equatorial plane for models RS1 -- RS4 at $t_{\rm c}$ approximately $3000-4000$\,yr.
Elapsed time after cloud begins to collapse $t$ and that after protostar formation $t_{\rm c}$ and protostellar mass $M_{\rm ps}$ (models RS2 and RS4) are given in the upper part of each panel.
Model name, sink condition and whether Ohmic dissipation included are given in each panel. 
}
\label{fig:14}
\end{figure}

Figure~\ref{fig:14} shows the density and velocity distributions for models RS1 -- RS4. 
For model RS2,  we could not calculate the cloud evolution for a long duration without Ohmic dissipation but with a higher spatial resolution ($h=0.6$\,AU) because the timestep became very small.
Thus, the cloud structure at $t_{\rm c} \simeq 3000$\,yr is plotted for model RS2, whereas that at $t_{\rm c} \sim 4000$\,yr is plotted for the other models.

Model RS1 has almost the same numerical settings as those in \citet{seifried12}, in which the sink accretion radius of $r_{\rm acc}=12.6$\,AU and threshold density of $n_{\rm thr}=3\times10^{11}\cm$ were adopted with a finest spatial resolution of $h=4.7$\,AU.
In addition, we did not impose Ohmic dissipation for this model. 
Note that the thermal evolution differs between our model and that of \citet{seifried12} because they considered the radiative cooling effect in more detail. 
We could not confirm any sign of disk formation in model RS1 by the end of the calculation, as seen in Figure~\ref{fig:14}{\it a}, which is consistent with \citet{seifried12}.
They claimed that disk formation was suppressed at $\mu \lesssim 10 $.

Model RS2 has a sufficiently high spatial resolution, but Ohmic dissipation was not imposed.
We also could not confirm the formation of a rotation-supported disk in this model. 
Although a small disk-like structure $\sim2$\,AU in size appears at the centre of the cloud (Fig.~\ref{fig:14}{\it b}), it is not supported by rotation.
Ohmic dissipation seems to assist disk formation even in massive clouds.
A rotation-supported disk appears during the early main accretion phase in models RS3 and RS4, in which Ohmic dissipation was imposed.
Although the difference in the sink accretion radius and threshold density between models RS3 and RS4 makes a small quantitative difference, almost the same disk appears in models RS3 and RS4, as shown in Figure~\ref{fig:14}{\it c} and {\it d}. 
Thus, both a higher spatial resolution ($h<1$\,AU) and Ohmic dissipation are necessary for investigating disk formation.

\begin{figure}
\includegraphics[width=150mm]{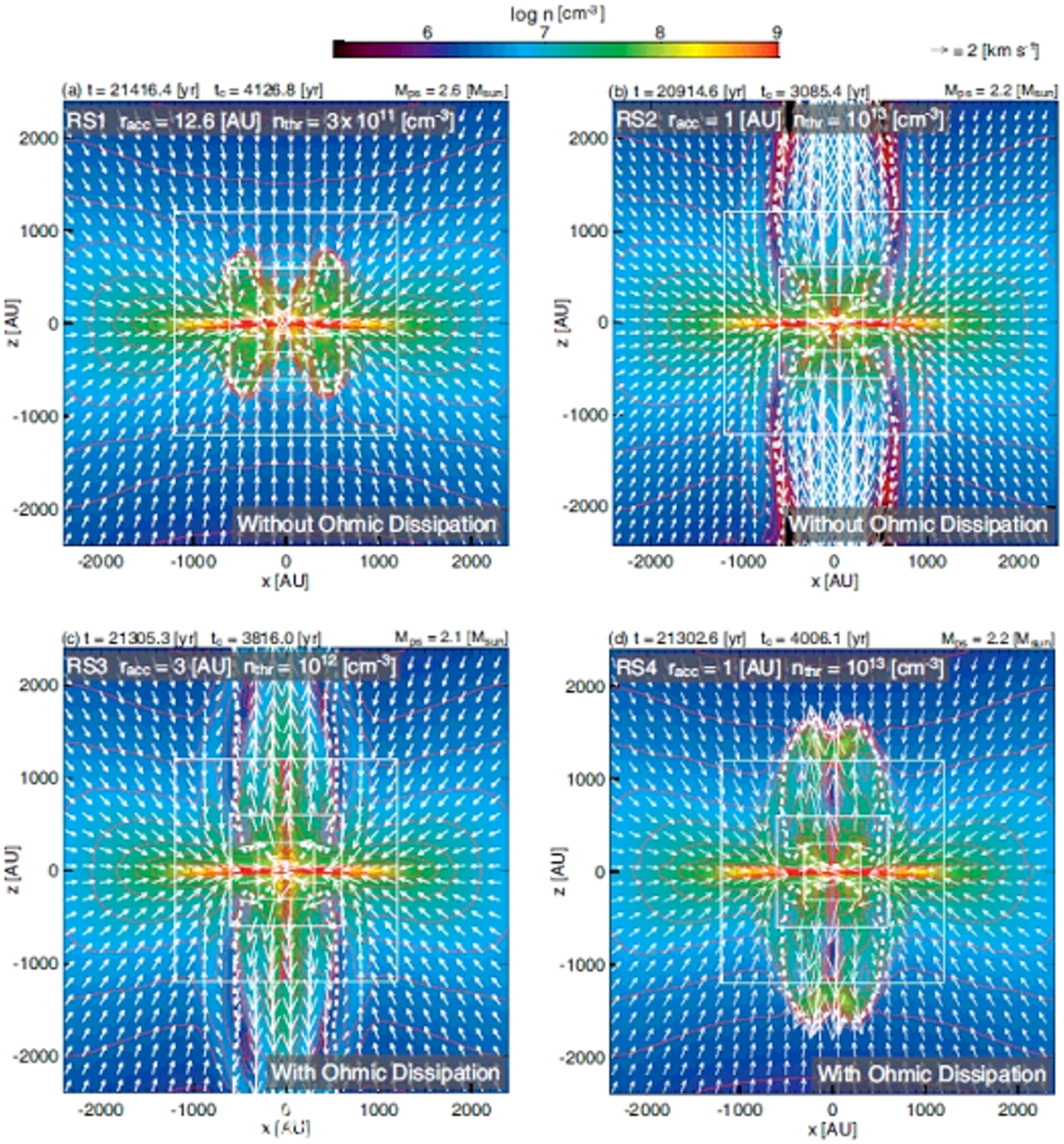}
\caption{
Same as  Fig.~\ref{fig:14}, but on the $y=0$ plane.
Box scale differs from that in Fig.~\ref{fig:14}.
White dotted line in each panel corresponds to the boundary between infalling and outflowing gas, inside of which the gas is outflowing from the centre of the cloud with a velocity exceeding sound speed (i.e. $v_r > c_s$).
}
\label{fig:15}
\end{figure}

Figure~\ref{fig:15} indicates that the difference in spatial resolution and sink treatment also affects the large-scale structure of the collapsing cloud.
A very weak outflow appears in model RS1, whereas a strong outflow is driven near the centre of the cloud for the other models.
Model RS2 has a maximum outflow speed of $>20$\,km\,s$^{-1}$. 
This is because the small core has a deeper gravitational potential and can drive the high-speed outflow.
In addition, the outflow speed and outflow momentum in model RS3 are larger than those in model RS4, and the sink accretion radius in model RS3 is larger than that in model RS4. 
The magnetic field dissipates effectively with a higher threshold density (or smaller sink accretion radius), because the magnetic dissipation region is limited to the high-density gas region \citep{nakano02}.
Thus, a relatively weak magnetic field is realized around the centre of the cloud with a higher threshold density, and a relatively weak outflow appears.
This indicates that the relation between the sink and outflow driving is not simple.

\begin{figure}
\includegraphics[width=150mm]{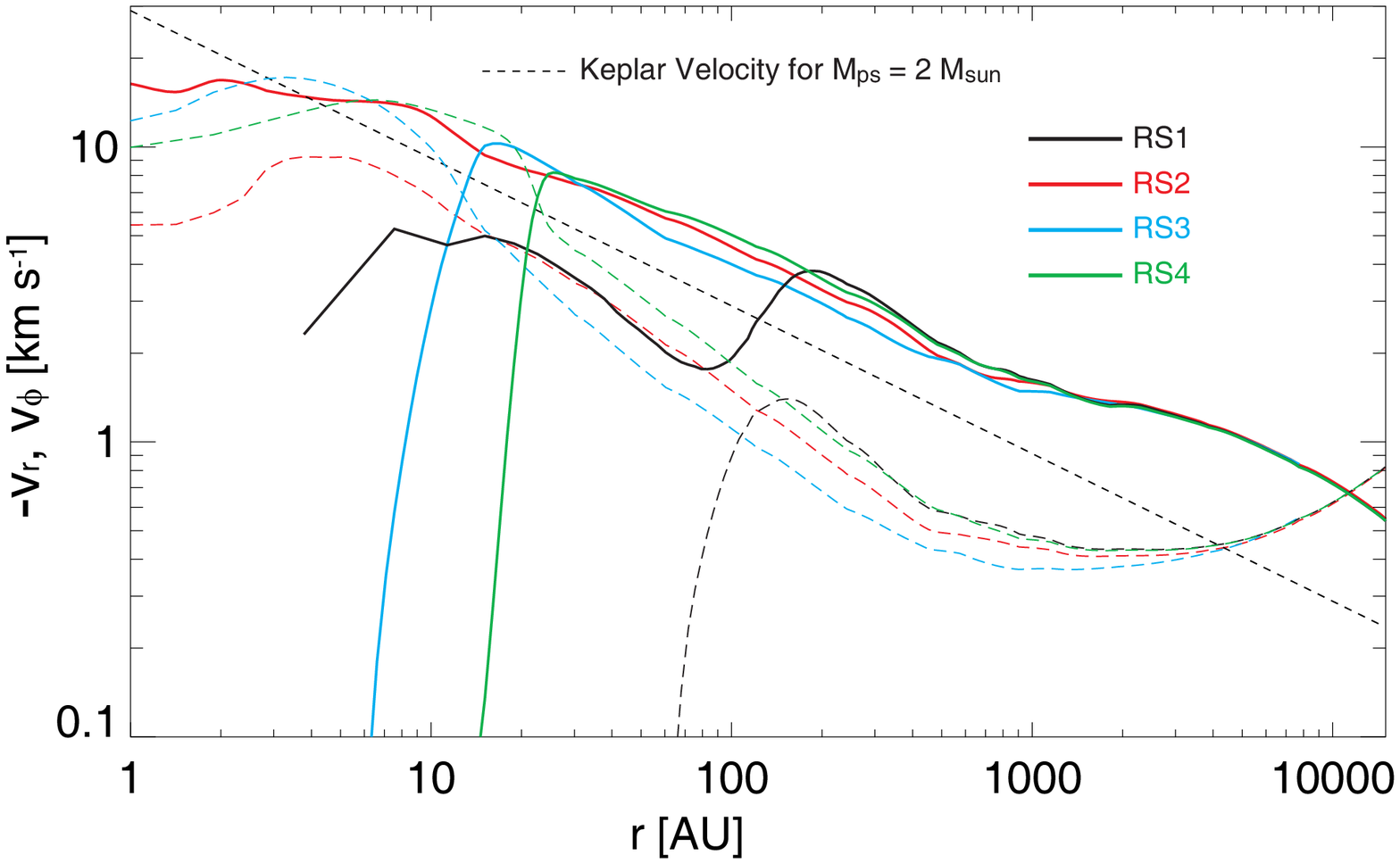}
\caption{
Radial $-v_r$ (solid line) and azimuthal $v_\phi$ (dashed line) velocities at the same epoch as in Figs.~\ref{fig:14} and \ref{fig:15} against the distance from the centre of the cloud for models RS1 -- RS4.
Keplerian velocity for $M_{\rm ps}=2\msun$ is also plotted (dotted line).
}
\label{fig:16}
\end{figure}

The (negative) radial and azimuthal velocities for models RS1 -- RS4 at the same epochs as in Figures~\ref{fig:14} and \ref{fig:15} are plotted in Figure~\ref{fig:16}; the Keplerian velocity for $M_{\rm ps}=2\msun$ is also plotted.
Both the radial and azimuthal velocities drop at $\sim100$\,AU in model RS1, which has a sink accretion radius of $r_{\rm acc} = 12.6$\,AU.
The velocity distribution in model RS1 differs considerably from those in the other models. 
In particular, the drop in the azimuthal velocity indicates an artificial effect of the larger accretion radius.
It is very difficult to estimate the mechanism of angular momentum transfer in a magnetized collapsing cloud.
However, the difference between the models without Ohmic dissipation (RS1 and RS2) implies that the angular momentum is artificially removed from the computational domain with a lower spatial resolution.

The figure also indicates that the disks in models RS3 and RS4 are supported by rotation because the azimuthal velocity is comparable to or exceeds the Keplerian velocity in the range of 3\,AU $\lesssim r \lesssim$\, 20\,AU. 
In addition, a smaller sink accretion radius and  higher threshold density form a larger disk: the disk sizes in models RS3 and RS4 are $\sim10$\,AU and $\sim 20$\,AU, respectively, when the protostellar mass reaches $M_{\rm ps}\sim 2\msun$. 
As a result, we have to carefully introduce the sink even in a massive star-forming cloud.

\section{Discussion}
\subsection{Spatial Resolution for Disk Formation}
\label{sec:resolution}
As described in \S\ref{sec:results}, different sink treatments yield qualitatively different outcomes even when the same initial conditions are adopted. 
A disk can form with both a smaller sink accretion radius and a higher threshold density, whereas no disk appears with a larger sink accretion radius.
Furthermore, different spatial resolutions seem to affect disk formation.
To investigate the effect of the spatial resolution on disk formation, we prepared the same initial cloud as in models US1 -- US8. 
Then, we calculated its evolution while changing the finest spatial resolution.
We call the model described in this subsection the low-resolution model (model USL, see Table~\ref{table:1}). 
For this model, we imposed the same sink properties as for model US3 ($r_{\rm acc}=6.7$\,AU and a variable threshold density).
However, different resolutions of the finest grid (or different levels of the maximum grid) were adopted.
Model US3 has a spatial resolution of $h=0.4$\,AU (Table~\ref{table:1}), whereas the low-resolution model has $h=1.6$\,AU.
Thus, around the centre of the collapsing cloud, the low-resolution model has four times coarser spatial resolution than model US3.

\begin{figure}
\includegraphics[width=150mm]{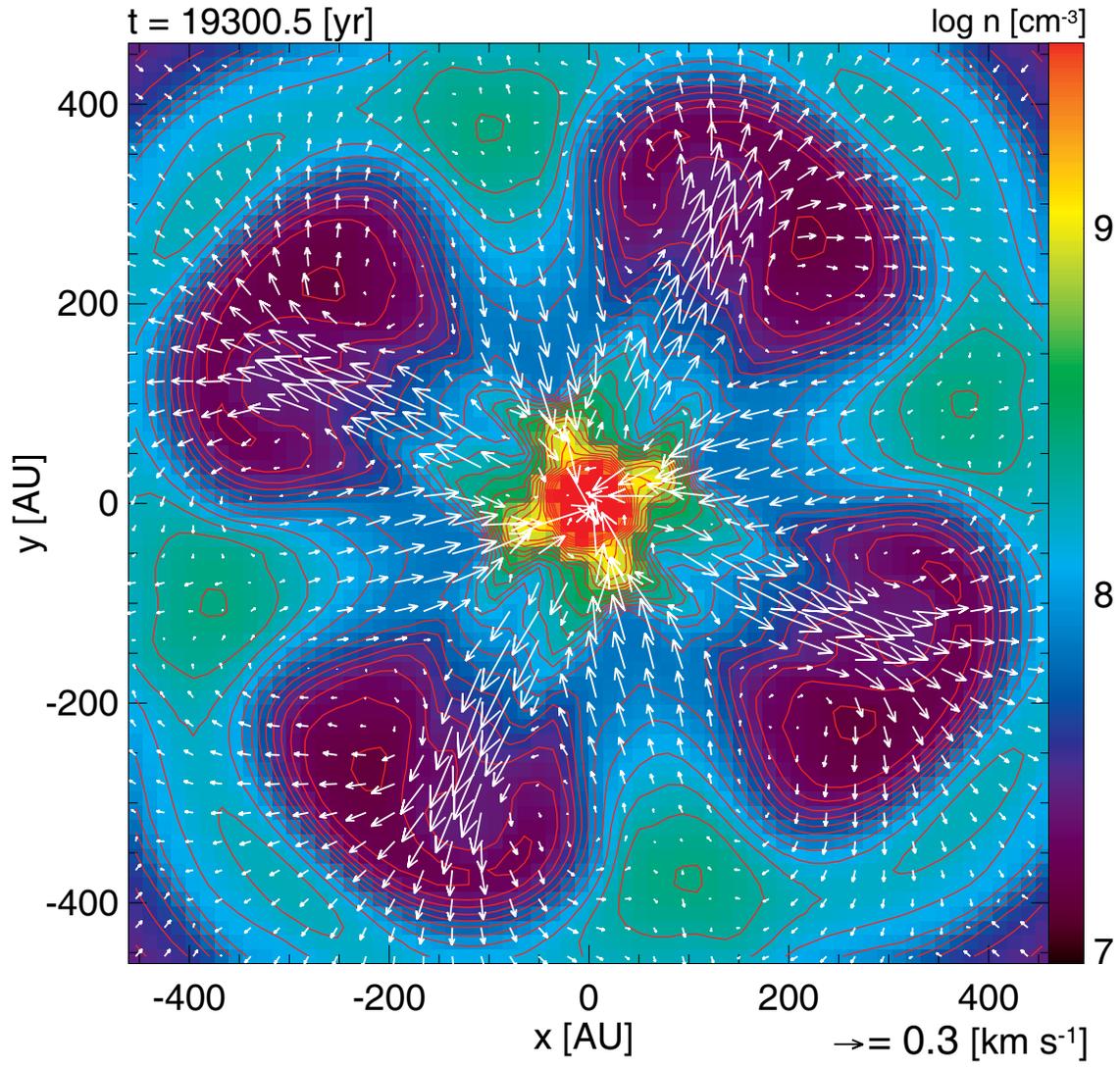}
\caption{
Density (colour and contours) and velocity distributions (arrows) on the equatorial plane for the low-resolution model.
Elapsed time after the cloud begins to collapse is given.
}
\label{fig:17}
\end{figure}

Figure~\ref{fig:17} shows the density and velocity distributions on the equatorial plane for the low-resolution model. 
Non-axisymmetric equatorial outflow and four cavities appear. 
Almost the same features are confirmed in some models of \citet{krasnopolsky11}, \citet{seifried12} and \citet{joos12}. 
In addition, model RJ1 shows the same features as shown in Figure~\ref{fig:11}{\it a}. 
This structure is thought to be caused by numerical reconnection \citep{li11} or interchange instability \citep{spruit90,spruit95,li96}. 
\citet{krasnopolsky10} pointed out that numerical (or artificial) reconnection suppresses disk formation during the early accretion phase. 
Because exploring this process is outside of the scope of this paper, we do not comment further on the mechanism.
However, we stress that a slight difference in the spatial resolution can make a great difference in the outcome.

In both US3 and the low-resolution model, no disk forms before the end of the calculation.
However, gas flows into the protostar, maintaining axisymmetry in model US3, as shown in Figure~\ref{fig:1}, whereas a complex structure appears in the low-resolution model. 
It is natural that numerical reconnection tends to occur with a lower spatial resolution.
Instead, interchange instability physically occurs in the collapsing cloud and may suppress disk formation. 
Moreover, in the same initial cloud, the disk forms with a much higher spatial resolution of $h\le0.2$ (models US7 and US8). 
It is expected that the disk can alleviate the conditions of interchange instability because the magnetic field is gradually distributed over the disk, not concentrated onto the sink.
We need further careful calculations to understand the conditions for disk formation, and the relation between the numerical resolution and reconnection and interchange instability.

\subsection{Outer Boundary Condition}
In addition to the spatial resolution and inner boundary condition (or sink condition), the outer boundary condition should affect  disk formation.
The gas above and below the disk is connected to the disk through the magnetic field lines and can brake the disk.
Thus, the outer boundary condition is very important for correct calculation of disk formation because the boundary matter can slow the disk rotation.
The angular momentum transferred by magnetic braking is stored in the infalling envelope when the star-forming cloud is assumed to be isolated in a very low-density interstellar medium. 
The stored angular momentum finally falls onto the centre of the collapsing cloud and forms a rotating disk. 
When the outflow appears near the protostar, it can escape from the star forming-cloud with an angular momentum.
Thus, only an outflow can really transfer the angular momentum into interstellar space.
Therefore, without angular momentum transfer by the outflow, the angular momentum should be conserved during the calculation or in the star-forming cloud, and a rotation supported disk forms as the infalling envelope dissipates.

To avoid the artificial effect of the outer boundary, we imposed it sufficiently far from the star-forming cloud, as described in \S\ref{sec:model}. 
However, we could not calculate the cloud evolution until almost all the infalling gas falls onto the centre of the cloud in this study because a  high spatial resolution was adopted to investigate the effect of the inner boundary condition (or sink).
Instead, \citet{machida11a} showed that a sufficiently large disk forms in the later accretion phase even when only a small disk exists in the early main accretion phase.
On the other hand, a rotation-supported disk never forms even during the later main accretion phase in \citet{li11}, in which the outflow also does not appear.
The outer boundary was imposed just outside the star forming cloud in \citet{li11}.
In this case, the angular momentum can be artificially canceled out at the outer boundary, and disk formation is  suppressed.
Moreover, if the mass inflows from the outer boundary, the infalling envelope never dissipates and continues to brake the disk through the magnetic field lines. 
Disk formation is also artificially suppressed in this case.
Therefore, we have to impose the outer boundary (or outer boundary condition) very carefully to investigate  disk formation.

\subsection{Necessary Condition for Disk Formation}
\label{sec:condition}
The calculation results seem to indicate that a finer spatial resolution is necessary for studying the formation process of the circumstellar disk. 
In principle, we should resolve the protostar itself to calculate disk formation because the disk forms near the protostar.
The protostar has a size of $\lesssim 0.01$\,AU and a number density of $\gtrsim 10^{20}\cm$.
Thus, a calculation timestep of $dt \lesssim 0.01$\,yr ($\equiv t_{\rm ff} \sim [G\rho]^{-1/2}$) is required with a spatial resolution of $h \lesssim 0.01$\,AU to resolve the protostar.
With this shorter timestep or a finer spatial resolution, we cannot calculate the evolution of the cloud and disk for a long duration after protostar formation. 
Instead, the sink can drastically promote the calculation (or time integration) and make it possible to investigate disk formation during the main accretion phase. 
Because the calculation timestep becomes longer with a larger sink accretion radius, we can calculate disk formation for a long duration with the sink.
However, a sink having a larger accretion radius seems to give a misleading result, as described in \S\ref{sec:results}.
Thus, we should determine the necessary conditions for the sink to adequately resolve the disk.
In other words, we should physically provide the sink accretion radius and/or threshold density to correctly calculate disk formation.
Resolving the first adiabatic core \citep{larson69,masunaga00} may provide a criterion for the sink for disk formation.

\begin{figure}
\includegraphics[width=150mm]{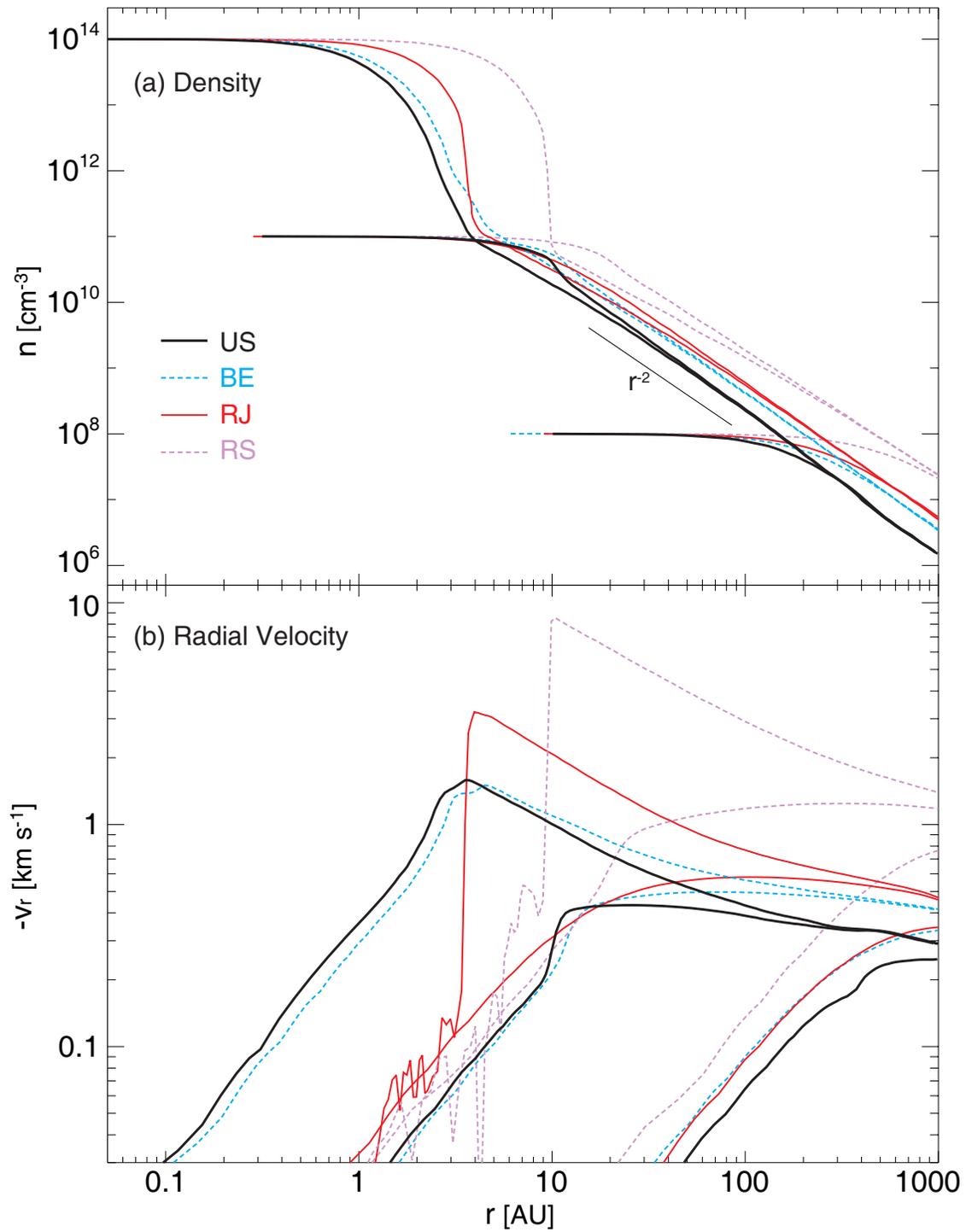}
\caption{
Density (upper) and velocity (lower) distributions at three different epochs ($n_{\rm c}=10^{8}\cm$, $10^{11}\cm$ and $10^{14}\cm$) of the gas collapsing phase for unmagnetized and non-rotating models (US, BE, RJ and RS).
The relationship $\rho \propto r^{-2}$ is also plotted in the upper panel.
}
\label{fig:18}
\end{figure}

To overview the evolution  of the collapsing cloud, we calculated the cloud evolution until the central density reaches $n=10^{15}\cm$ above which molecular hydrogen begins to dissociate and a protostar (or a second core) forms \citep{larson69,masunaga00}. 
Because it is difficult to quantitatively compare the cloud evolution among models with rotation and a magnetic field, we prepared  unmagnetized ($B_0 = 0$\,$\mu$\,G) and non-rotating ($\Omega_0=0$\,s$^{-1}$) clouds that have the same initial density distributions as the models listed in Table~\ref{table:1}.
Figure~\ref{fig:18} plots the density (top panel) and velocity (bottom panel) distributions  against the distance from the centre of the cloud for the US, BE, RJ and RS models when the central density reaches $n=10^8\cm$, $10^{11}\cm$ and $10^{14}\cm$, respectively, where the models have the same initial density distribution as in models US1 -- US8 (US), BE1 - BE6 (BE), RJ1 - RJ4 (RJ) and RS1 -- RS4 (RS). 
Although there are slight quantitative differences, the density and velocity distributions are qualitatively the same as those in \citet{larson69} and \citet{masunaga00}.
Before the central density reaches $n_{\rm c} \sim 10^{11}\cm$, the density distribution obeys $\rho \propto r^{-2}$ in the outer envelope and $\rho\propto r^0$ in the inner region.
When the central density reaches $n\sim10^{11}\cm$, a shock occurs and the first (adiabatic) core forms.
As shown in the figure, the first core is $r=2-10$\,AU in size at its formation. 
Inside the first core, the radial velocity quickly slows to $-v_r<0.1$\,km\,s$^{-1}$.

\citet{saigo06} showed that the first core (or the remnant of the first core) with rotation is sustained for at least $\sim 10^3-10^4$\,yr \citep[see also][]{tomida10}. 
\cite{bate98,bate10} pointed out that the first core is the origin of the circumstellar (or rotation-supported) disk, indicating that the circumstellar disk has a typical size of $1-3$\,AU at its formation \citep[see also][]{machida10a}. 
Thus, we require a spatial resolution of at least $1-3$\,AU to investigate disk formation.
Although they calculated the evolution of the unmagnetized cloud, the first core is also expected to play an important role in disk formation in magnetized clouds.

\begin{figure}
\includegraphics[width=150mm]{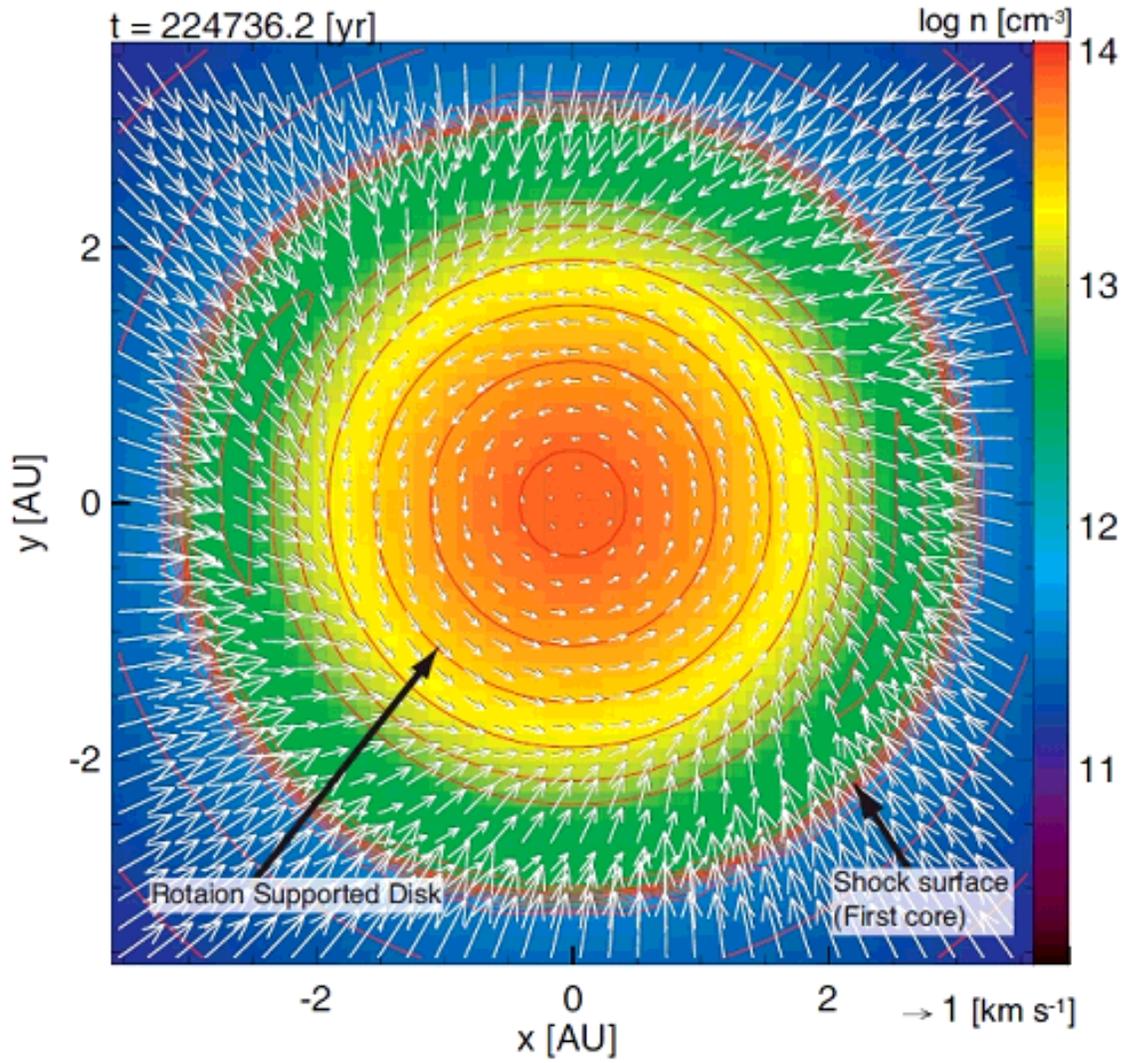}
\caption{
Density (colour and contours) and velocity (arrows) distributions on the equatorial plane for model BEH before protostar formation.
Black arrows indicates the rotation-supported disk and shock surface.
}
\label{fig:19}
\end{figure}

To investigate disk formation in the very early phase of star formation and the relation between the first core and the circumstellar disk, we prepared model BEH, which has the same initial conditions as model BE3 but a different spatial resolution ($h=0.06$\,AU) and sink properties ($ r_{\rm acc}=0.1$\,AU and $n_{\rm thr}=10^{14}\cm$; see Table~\ref{table:1}).
Note that with a higher spatial resolution, we could not calculate the cloud evolution for a longer duration (more than $\sim1000$\,yr after protostar formation) for model BEH. 
Figure~\ref{fig:19} shows the density and velocity distributions just before protostar formation for model BEH. 
Before protostar formation, the first core appears surrounded by the shock.
The size of the first core in model BEH is $\sim3$\,AU, which is comparable to that in the unmagnetized and non-rotating model (Fig.~\ref{fig:18}).
The radial velocity of the gas significantly decreases after it passes through the shock front (or the surface of the first core), because the thermal pressure gradient force impedes further collapse inside the first core.

In an unmagnetized cloud, the collapse timescale becomes longer than the rotation timescale, and  the azimuthal velocity dominates the radial velocity (i.e. $v_{\phi} > v_r$) inside the first core.
Therefore, a rotation-supported disk appears \citep{bate98,bate10,bate11,saigo06,saigo08,machida10a}.
On the other hand, in a magnetized cloud, the gas can collapse slowly even inside the first core as shown in Figure~\ref{fig:19}, because the angular momentum is transferred by the magnetic field. 
However, a longer collapse timescale allows  sufficient time for dissipation of the magnetic field. 
Because the first core has a density of $n\gtrsim 10^{11}\cm$ at its formation (Fig.~\ref{fig:18}), the magnetic field can dissipate by Ohmic dissipation \citep{nakano02}.
As a result, the magnetic field dissipates significantly inside the first core, and magnetic braking becomes ineffective. 
Therefore, a rotation-supported disk forms inside the first core, as shown in Figure~\ref{fig:19}.
The rotation-supported disk is $\sim1-2$\,AU in size at its formation.
It is natural that with the first core unresolved, no disk appears in the magnetized cloud, as shown in Figures~\ref{fig:1} and \ref{fig:3}, because there is not enough time for dissipation of the magnetic field.

\begin{figure}
\includegraphics[width=150mm]{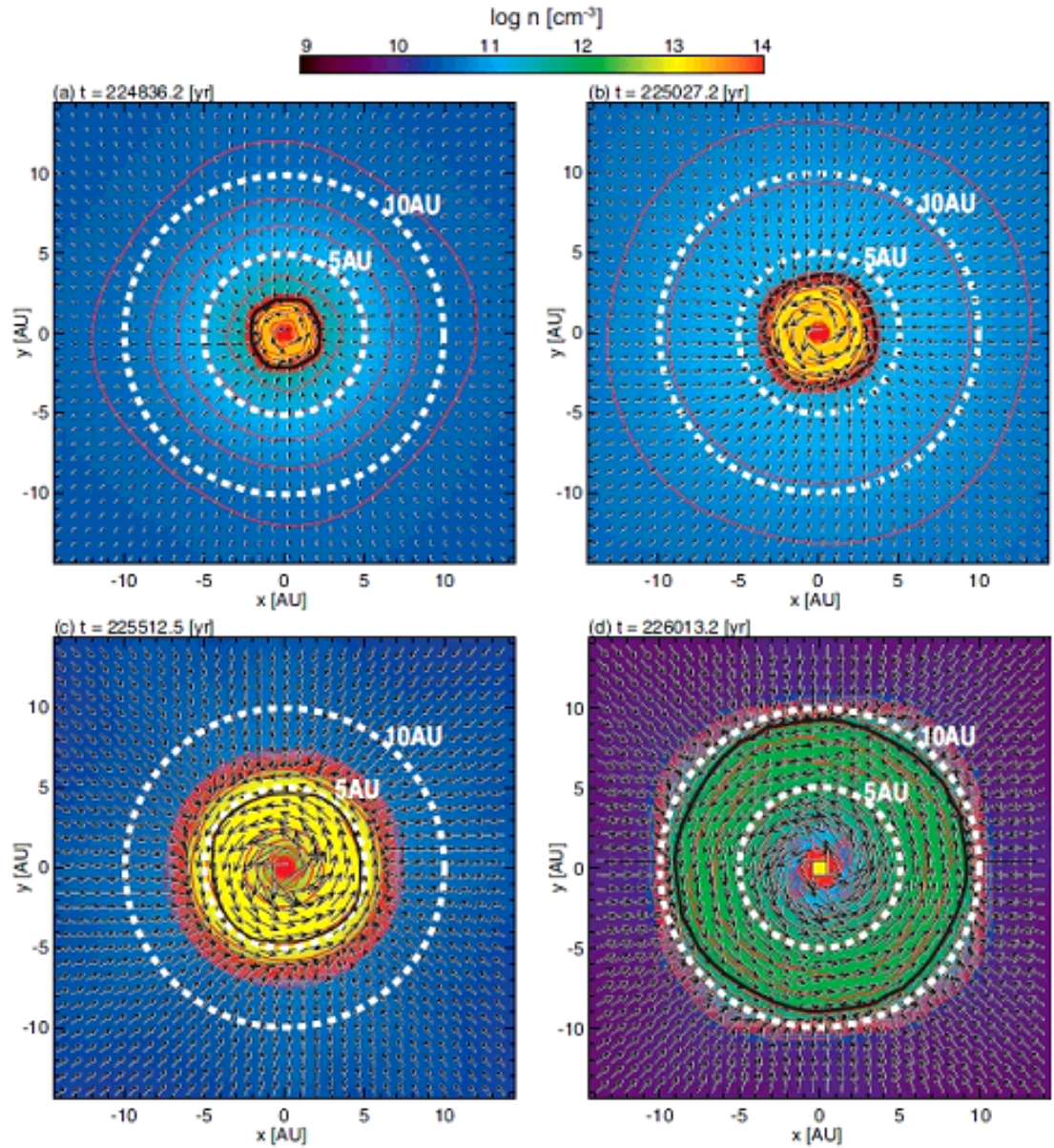}
\caption{
Density (colour and contours) and velocity (arrows) distributions on the equatorial plane for model BEH during the early main accretion phase.
Elapsed time after the cloud begins to collapse $t$ is given  in each panel. 
White dotted circle indicate radii of $5$\,AU and $10$\,AU.
Black circle marks the contour of $n=10^{12}\cm$, inside of which the magnetic field is significantly dissipated by Ohmic dissipation.
}
\label{fig:20}
\end{figure}

After the rotation-supported disk forms inside the first core, the collapsing gas is supported by rotation and gas accumulates in the disk.
Therefore, the high-density gas region, which corresponds to the disk, increases. 
Because the magnetic dissipation region corresponds to the high density region, the magnetic field effectively dissipates in this region.
Therefore, the disk expands further as the magnetic dissipation region increases.
The evolution of the rotation-supported disk during the early main accretion phase is plotted in Figure~\ref{fig:20}.
The figure indicates that the disk has is $\sim 2$\,AU in size just after protostar formation (Fig.~\ref{fig:20}{\it a}).
Then, it gradually expands with time and exceeds $10$\,AU about $1000$\,yr after  protostar formation (Fig.~\ref{fig:20}{\it d}).
In addition, the disk, which is surrounded by the shock (see red contours), has a density of $>10^{12}\cm$ (black contour line) at each epoch. 
The magnetic field dissipates significantly in the region of $n>10^{12}\cm$. 
The figure also indicates that a spatial resolution of $\lesssim 1-2$\,AU is necessary to calculate disk formation.

\begin{figure}
\includegraphics[width=150mm]{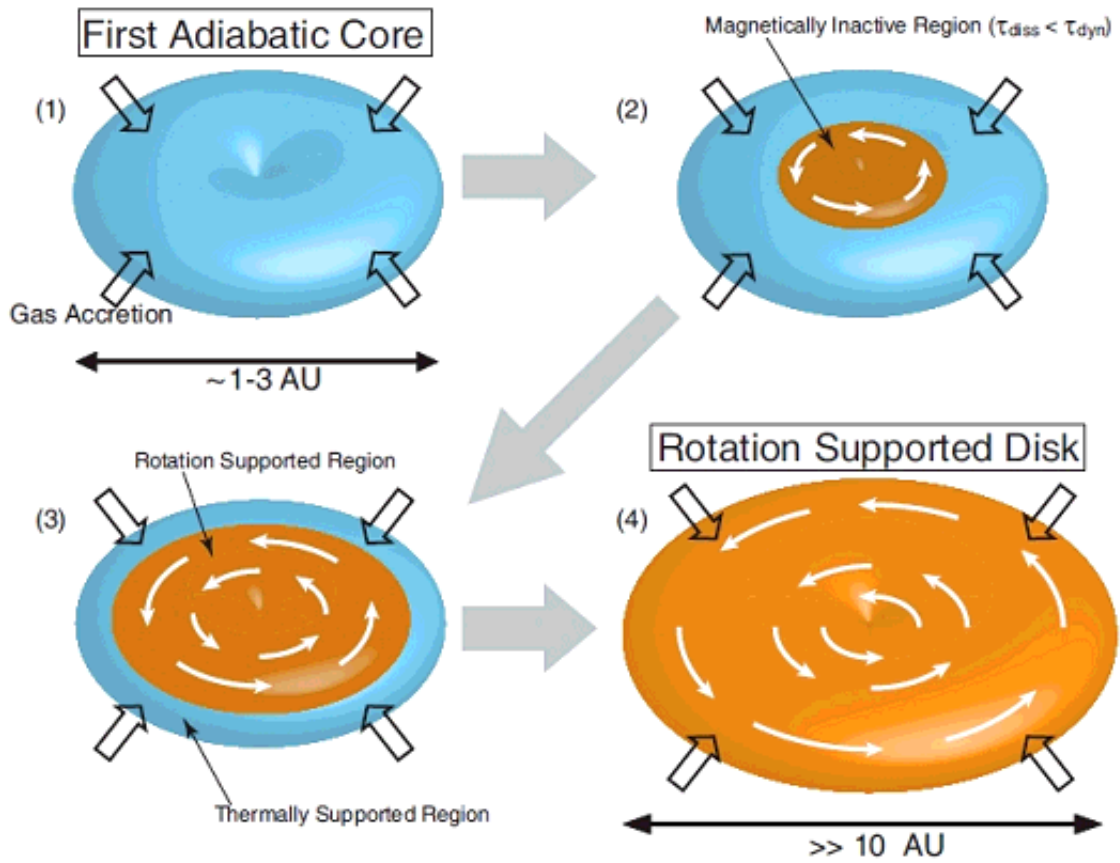}
\caption{
Schematic view of disk formation.
}
\label{fig:21}
\end{figure}

The formation process of the circumstellar disk is schematically summarized in Figure~\ref{fig:21}.
The first core forms before protostar formation (Fig.~\ref{fig:21}, epoch [1]).
Because the first core has a density of $n\gtrsim 10^{12}\cm$ and the gas collapses very slowly inside the first core, the magnetic field effectively dissipates inside it (Fig.~\ref{fig:21}, epoch [2]).
Then, the rotation-supported region appears in the first core (Fig.~\ref{fig:21}, epoch [3]), as also seen in Figure~\ref{fig:19}.
At this epoch, the disk-like structure is partly supported by thermal pressure.
Finally, the magnetically inactive region (or the high-density region), which is supported by rotation,  expands and the rotation-supported disk forms (Fig.~\ref{fig:21} epoch [4]).

\subsection{Dependence of Initial Condition on Disk Formation}
The initial conditions of the cloud also affect disk formation.
We prepared four types of initial cloud having different density distributions;  the mass-to-flux ratio is adjusted to so that $\mu \simeq 3$ in each model. 
In addition, the rotation parameter is almost the same among the models ($\beta_0 = 0.02-0.03$).
Note that massive cloud models RS1 -- RS4 have slightly different cloud parameters to match a typical model in \citet{seifried12}.
Although the initial clouds have almost the same parameters, the resultant disk size differs among the models.
The disk has a size of $\sim 20$\,AU when the protostellar mass reaches $M_{\rm ps} \simeq 0.5\msun$ in the uniform density model (US8, see Fig.~\ref{fig:3}), whereas it is $\sim 50-100$\,AU in size at the same epoch in the Bonnor-Ebert density models (BE1 -- BE6, see Fig.~\ref{fig:7}).

In addition, for disk formation, different sink conditions seem to be required for models having initially different density distributions.
In the uniform density models, a rotation supported disk never appears with a lower sink threshold  density ($n_{\rm thr}\lesssim 10^{12}\cm$) and/or a larger sink accretion radius ($r_{\rm acc}\gtrsim 1$\,AU).
In Figure~\ref{fig:3}, no clear disk appears with $n_{\rm thr}< 10^{13}\cm$, whereas a rotating disk forms with $n_{\rm thr}=10^{14}\cm$.
On the other hand, a rotating disk appears with a lower threshold density and larger sink accretion radius in the Bonnor-Ebert  models (Fig.~\ref{fig:7}).
For a variable threshold density and an accretion radius of $r_{\rm acc}=3$\,AU, a rotation-supported disk forms in Bonnor-Ebert  models BE1, BE2, BE3, BE5 and BE6.
Instead, with the same sink condition, no disk appears in the uniform density model (US 2). 
Moreover, a rotation-supported disk appears without a sink but with Ohmic dissipation in model RJ3.
These results indicate that the sink conditions for disk formation differ in models with initially different density distributions.

We could not find any obvious reason that a different initial density distribution yields different disk sizes and sink conditions.
We expect that one possible reason is the size of the first core.
In the collapsing cloud, the density and velocity obey the self-similar solution \citep{larson69,masunaga00}. 
Thus, the collapsing cloud should have almost the same density and velocity profiles.
However, they slightly differ: the first core is smaller for in the uniform density models than in the other models, as shown in  Figure~\ref{fig:18}. 
As described in \S\ref{sec:condition}, we need to resolve the first core to correctly calculate disk formation. 
Although the size differences in the first core are not significant among the models, Figure~\ref{fig:18} indicates that  a higher spatial resolution is required for the uniform density models than for the other models.

Another possibility is the initial distribution of the mass-to-flux ratio and specific angular momentum.
When the clouds have the same mass-to-flux ratio and a uniform magnetic field is assumed, the magnetic field in the uniform density models is stronger than that in the non-uniform density models.
For example, model US1 has $\mu=2.9$ and $B_0 = 35\,\mu$\,G, whereas model BE1 has $\mu=3$ and $B_0=14.3$. 
Thus, the magnetic field for US1 is 2.5 times stronger than that for model BE1, although both clouds have almost the same (central) density of $\sim10^5\cm$. 
In addition, the mass-to-flux ratio is a function of the radius and increases with the radius around the centre of the cloud, as shown in \citet{machida11b}. 
In the outer cloud region, the mass-to-flux ratio decreases with the radius in the Bonnor-Ebert models, whereas it continues to increase in the uniform density models.
As a result, with the same mass-to-flux ratio, the central magnetic field for the Bonnor-Ebert model is weaker than for the uniform density model. 
Because only the gas around the centre of the cloud contributes to disk formation in very early  of star and disk formation, the distribution of the mass-to-flux ratio and density should affect the magnetic braking.

\subsection{Thermal Evolution and Equation of State}
The first core is crucial to studies of disk formation.
As described in \S\ref{sec:condition}, a rotation supported disk appears after the gas collapse temporarily halts or slows down and the magnetic field dissipates inside the first core. 
At its formation, the first core is supported mainly by thermal pressure and partially by rotation and the magnetic field. 
Thus, to correctly calculate the first core and its subsequent evolution, we should investigate the cloud evolution using radiation MHD (RMHD) simulations. 
However, such calculations require huge CPU resources and it is very difficult to use them to investigate disk formation. 
Instead of solving the energy equation and calculating the radiative transfer, previous studies used a barotropic equation of state to realize  a long-term calculation possible \citep{mellon08, mellon09, hennebelle09, dapp10, dapp12, li11, krasnopolsky10,  krasnopolsky12, machida11b, joos12}.
The barotropic equation of state mimics the thermal evolution of the collapsing cloud calculated by a radiation hydrodynamics (RHD) simulation \citep{larson69,masunaga00,commerson10,tomida12}. 
The RHD calculations showed that the collapsing gas behaves isothermally as $P\propto \rho $ for $\rho<\rho_{\rm cri}$, whereas it behaves adiabatically as $P\propto \rho^\gamma $ for $\rho > \rho_{\rm cri}$, where $\rho_{\rm cri}\simeq 10^{-13}$\,g$\cm$.

As the polytropic index, $\gamma=7/5$ was used in \citet{mellon08,mellon09}, \citet{hennebelle09}, \citet{li11}, \citet{krasnopolsky10,  krasnopolsky12} and \citet{joos12}, whereas $\gamma=5/3$ was adopted in \citet{machida11b}.
\citet{tomida12} showed that the collapsing gas has a polytropic index of $\gamma \simeq  5/3$ in the range of $10^{-13}$\,g$\cm$ $\lesssim \rho \lesssim $ $10^{-11}$\,g$\cm$ ($3\times 10^{10}\cm  \lesssim n \lesssim  3\times 10^{12}\cm$), whereas it has $\gamma \simeq 7/5$ in the range of $10^{-11}$\,g$\cm$ $\lesssim \rho \lesssim $ $10^{-8}$\,g$\cm$ ($3\times 10^{12}\cm  \lesssim n \lesssim  3\times 10^{15}\cm$).
With sink, the high-density gas at $n\gtrsim 10^{12}-10^{13}\cm$ is removed from the computational domain (\S\ref{sec:sink}).
Therefore, only the gas having $n < 10^{12}-10^{13}\cm$ remains in the computational domain.
Thus, we should use not $\gamma=7/5$ but rather use $\gamma=5/3$ as the polytropic index.
Alternatively, we should further divide the adiabatic equation of state in the range of $10^{-13}$\,g$\cm$ $\lesssim \rho \lesssim $ $10^{-8}$\,g$\cm$, as done in  \citet{dapp12}.

In this study, to investigate the effect of the initial cloud configuration and sink on the disk formation and compare our results with previous studies, we purposely used $\gamma = 7/5$.
However, with a hard equation of state ($\gamma=5/3$), the adiabatic core (first core) is expected to persist for a long duration.
The magnetic field dissipates and a rotation-supported disk forms inside the adiabatic core, as described in \S\ref{sec:condition}.
Thus, a long-lived first core with a harder equation of state is thought to assist the formation of a rotation-supported disk. 
Note that during the (later) main accretion phase, heating from the protostar and radiative cooling should affect the thermal evolution around the protostar.
Thus, in principle, we should calculate disk formation using RMHD simulations.

\section{Summary}
In this study, we investigated the effects of the initial cloud configuration and sink on  disk formation in a magnetized collapsing cloud. 
We prepared four different initial clouds that have almost the same mass-to-flux ratio and ratio of  rotational to gravitational energy. 
Each model is identical to a fiducial model in \citet{li11}, \citet{machida11b}, \citet{joos12} or \citet{seifried12}, in which the same density distribution of the initial cloud as in previous studies was adopted.
In addition, we adopted 15 different sink treatments in combination with  different sink accretion radii $r_{\rm acc}$ and threshold densities $n_{\rm thr}$; the cell width of the finest grid was also changed in each model to resolve the sink accretion radius. 
As a fiducial model of each cloud model, we adopted the same sink condition and spatial resolution as those in previous studies.

As a result of the calculations, we could reproduce any result shown in previous studies \citep{li11,machida11b,joos12,seifried12}. 
We confirmed that as claimed in previous studies, no disk forms in a strongly magnetized cloud having uniform \citep{li11} and non-uniform \citep{joos12,seifried12} density distributions with the same sink conditions adopted in the corresponding previous study. 
We also confirmed that even when no disk forms with the same sink conditions as in previous studies, a rotation-supported disk can form with slightly different sink conditions. 
The protostellar outflow is also strongly influenced by the sink.
We concluded that disk formation and outflow driving are very sensitive to the sink treatment, and the sink can reproduce any outcome.

In a strongly magnetized cloud, the sink condition necessary for disk formation differs among models that have different initial density distributions.
For a uniform sphere, both a considerably smaller sink accretion radius $r_{\rm acc} \lesssim 1$\,AU and a higher threshold density $n_{\rm thr}\gtrsim 10^{13}-10^{14}\cm$ seem to be required for investigating the formation of a rotation supported disk.
However, for the uniform density cloud, we could not confirm convergence of the calculation results such as the disk size and mass, and outflow momentum with a smaller sink radius and higher threshold density. 
It is difficult to calculate long-term evolution of the disk with a smaller sink accretion radius.
Thus, we are not sure whether the rotation-supported disk really forms in the uniform cloud  with $\mu\lesssim 3$.
In reality,  although rotation-supported disk appeared once with a relatively lower threshold density of $n_{\rm thr}=10^{13}\cm$, it suddenly disappeared during the main accretion phase, as shown in Figure~\ref{fig:3}{\it c} (model US7).
Although the rotation-supported disk remains  at the end of the calculation with a relatively higher sink threshold density of $n_{\rm thr}=10^{14}\cm$ (model US8), it may also disappear in a subsequent evolutionary stage. 
Thus, an even higher spatial resolution may be required for the uniform cloud to investigate disk formation \citep{dapp12}.

On the other hand, for clouds with a non-uniform density profile, a rotation-supported disk forms even with a larger sink accretion radius of $r_{\rm acc} \ge 3$\,AU or a lower threshold density of $n_{\rm thr}\le 10^{13}\cm$.  
For the non-uniform clouds, a necessary condition of the sink is $r_{\rm acc}\le 1- 3$\,AU and $n_{\rm thr}\ge10^{13}\cm$. 
With these conditions, we can safety calculate disk formation in a strongly magnetized cloud.
As a result, the sink conditions for disk formation are more severe in the uniform cloud than in the non-uniform cloud.
The difference in the sink conditions is expected to result from the size of the first core and/or initial distribution of the mass-to-flux ratio and specific angular momentum. 
Although we could not determine the sink conditions for the uniform cloud, we may not need to pursue it because the host cloud for star formation is not uniform.
Instead, we may have to investigate the effects of the initial distribution of the mass-to-flux ratio and specific angular momentum on disk formation in future studies.

The outflow is also influenced by the sink. 
In the same initial cloud,  no or very weak outflow appears with a larger sink accretion radius, whereas a powerful outflow tends to appear with a smaller sink accretion radius or higher threshold density. 
However, the outflow momentum or outflow speed does not necessarily increase with a higher spatial resolution or smaller sink accretion radius (Fig.~\ref{fig:15}).  
Ohmic dissipation becomes effective near the protostar and forms a magnetically inactive region by which the outflow cannot be driven.
Thus, when the high-density gas region is well resolved, the outflow momentum weakens somewhat. 
Although there is a slight quantitative difference, the outflow momentum for models with initially non-uniform density corresponds well to the observations as described in \S\ref{sec:results}.
On the other hand, the outflow momentum in the uniform cloud models is more than one order of magnitude smaller than the observations, which may indicate that the host cloud for star formation is not described by the uniform density profile.

Next, we describe the early formation of the rotation supported disk.
Previous studies investigating cloud evolution in an unmagnetized cloud showed that  the first core is the origin of the rotation supported disk \citep{bate98,bate10,bate11,walch09, machida10a,tsukamoto11,tsukamoto13}. 
In this study, we showed that the first core is a significant ingredient in disk formation even in magnetized clouds \citep[see also][]{dapp10,dapp12,inutsuka12}.
The first core is formed before protostar formation and is sustained for $\sim10^3-10^4$\,yr by rotation.
The radial velocity of the infalling gas quickly decreases as it passes through the shock of the first core, and thermal pressure prevents the gas from rapidly collapsing toward the protostar inside the first core (remnant). 
Because the first core has a density of $n\gtrsim10^{11}-10^{12} \cm$ and the gas stays inside it for a long duration, the magnetic field effectively dissipates.
Then, magnetic braking becomes ineffective, and a rotation-supported disk forms.
Thus, the initial size of the disk roughly corresponds to the size of the first core, $r\sim1$\,AU.
After the rotation-supported region appears, the rotating gas accumulates around it, and the magnetically inactive region expands.
Finally, the rotation-supported region is expanded outside the first core, and a rotation-supported disk forms (Fig.~\ref{fig:21}).
We conclude that the necessary condition for disk formation is to resolve the first core which has a size of $\sim1$\,AU and a density of $n\gtrsim10^{12}\cm$. 
Except for the uniform density models that is not a realistic initial condition for star-forming clouds, the rotation supported disk appears with magnetic dissipation  when the sink condition of $r_{\rm acc}\le1$\,AU and $n\ge10^{13}\cm$ is given. 
This condition corresponds well to the condition for resolving the first core, and is necessary for  disk formation. 
Note that a smaller first core appears in uniform density models with a strong magnetic field, a more high spatial resolution may be required for these models.

Finally, we comment on the subsequent evolution of the circumstellar disk.
In this study, with a higher spatial resolution, we could not calculate the disk evolution for a long duration. 
In many models, more than half of the initial cloud mass remains in the infalling envelope.
Because the gas that accretes later has a larger specific angular momentum, it contributes greatly to the further evolution of the disk.
In addition, the angular momentum transferred by magnetic braking is stored in the infalling envelope that finally accretes onto the centre of the cloud. 
Although part of the angular momentum in the star-forming cloud is expelled into interstellar space by the outflow, most of the angular momentum can contribute to the formation and evolution of the disk, especially in the later accretion phase.
In this study, the rotation-supported disk was only $\sim10$\,AU in size in some models.
However,  the disk is expected to grow exponentially in the later main accretion phase, as described in \citet{machida11b}.
To determine the final properties of the circumstellar disk, we need to calculate the formation and evolution of the disk for a longer time with appropriate sink conditions.

\section*{Acknowledgements}
We have benefited greatly from discussions with  ~K. Tomida. 
Some model settings were given by ~M. Joos.
Numerical computations were carried out on NEC SX-9 at Center for Computational Astrophysics, CfCA, of National Astronomical Observatory of Japan.
This work was supported by Grants-in-Aid from MEXT (23540270, 25400232).

\clearpage
\begin{table}
\setlength{\tabcolsep}{3.5pt}
\caption{Model parameters}
\label{table:1}
\begin{center}
\begin{tabular}{c|cccccccccc|ccccccc} \hline
{\footnotesize Model} & 
$n_{\rm c,0}$  & $M_{\rm c}$  &  
$r_{\rm c}$  & $B_0$  &  $\Omega_0$    
& \multirow{2}{*}{$\alpha_0$} & \multirow{2}{*}{$\beta_0$} & \multirow{2}{*}{$\gamma_0$} & \multirow{2}{*}{$\mu $} &
\multirow{2}{*}{Sink} & r$_{\rm acc}$ & n$_{\rm thr}$  & $h$  & \multirow{2}{*}{O.D.$^1$} & \multirow{2}{*}{RSD$^2$} & \multirow{2}{*}{Out$^3$} \\

 & 
 {\scriptsize [cm$^{-3}$]} &  {\scriptsize [$\msun$]} &  
 {\scriptsize [AU]}  & [$\mu$\,G] &  {\scriptsize   [$10^{-13}$\, s$^{-1}$]} &  
&  &  &  &   
 & {\scriptsize [AU]} & {\scriptsize [cm$^{-3}$]} & {\scriptsize [AU]} &  &  & \\
\hline

US1 & \multirow{9}{*}{$10^5$}  & \multirow{9}{*}{1.0} & \multirow{9}{*}{$6.7 \times 10^3$} & \multirow{9}{*}{35.4 (U)}  & \multirow{9}{*}{1} 
& \multirow{9}{*}{0.8} & \multirow{9}{*}{0.03} & \multirow{9}{*}{0.13} & \multirow{9}{*}{2.9} & Y & 1    & V  &  0.4 & Y &  N & N \\
US2 &  &  &  &  &  &   &  &  &  & Y & 3.4  & V   &  0.4 & Y &  N & N \\
US3 &  &  &  &  &  &   &  &  &  & Y & 6.7  & V   &  0.4 & Y &  N & N \\
US4 &  &  &  &  &  &   &  &  &  & Y & 13.4 & V   &  0.4 & Y &  N & N \\
US5 &  &  &  &  &  &   &  &  &  & Y &    1 & 10$^{11}$  & 0.4 & Y &  N & Y(?) \\
US6 &  &  &  &  &  &   &  &  & & Y &    1 & 10$^{12}$  & 0.4 & Y &  N & Y(?) \\
US7 &  &  &  &  &  &   &  &  &  & Y &    1 & 10$^{13}$  & 0.2 & Y &  Y & Y \\
US8 &  &  &  &  &  &   &  &  &  & Y &    1 & 10$^{14}$  & 0.1 & Y &  Y & Y \\
USL &  &  &  &  &  &   &  &  &  & Y & 6.7  & V   &  1.6 & Y &  N & N \\
\hline
BE1 & $1.4\times10^5$  & 2.1 & \multirow{7}{*}{$1.5\times10^4$} & 14.3 (U) &$0.81$ & 0.6 & \multirow{7}{*}{0.02} & 0.10 & 3.0 & Y  & 3 & V  & 0.5 & Y & Y & Y\\
BE2 & $1.7\times10^5$  & 2.6 & & 17.2 (U) &$0.87$ & 0.5 &  & 0.10 & 3.0 & Y  & 3 & V  & 0.5 & Y & Y & Y \\
BE3 & $2.1\times10^5$  & 3.2 & & 21.5 (U) &$0.98$ & 0.4&   & 0.10 & 3.0 & Y  & 3 & V  & 0.5 & Y & Y & Y \\
BE4 & $1.7\times10^5$  & 2.6 & & 17.2 (U) &$0.87$ & 0.5 &  & 0.10 & 3.0 & Y  & 1 & $10^{12}$  & 0.5 & Y & Y & Y \\
BE5 & $1.7\times10^5$  & 2.6 & & 37.8     &$0.87$ & 0.5 &  & 0.14 & 3.0 & Y  & 3 & V  & 0.5 & Y & Y & Y \\
BE6 & $1.7\times10^5$  & 2.6 & & 23.2 (U) &$0.87$ & 0.5 &  & 0.30 & 1.7 & Y  & 3 & V  & 0.5 & Y & Y & Y \\
BEH & $1.7\times10^5$  & 2.6 & & 17.2 (U) &$0.87$ & 0.5 &  & 0.10 & 3.0 & Y & 0.2 & $ 10^{14}$     & 0.06   & Y & Y & Y \\

\hline
RJ1 & \multirow{4}{*}{$8\times10^6$} & \multirow{4}{*}{1.0} & \multirow{4}{*}{$3\times10^3$} & \multirow{4}{*}{257 (U)}  & \multirow{4}{*}{$5$} & 
\multirow{4}{*}{0.4} & \multirow{4}{*}{0.03} & \multirow{4}{*}{0.21} & \multirow{4}{*}{3.0} & N & --- &  ---       & 0.4 & N & N(?) & Y \\
RJ2 &  &  &  &   &  &  &  &  &   & Y & 3   &  ---       & 0.4 & N & Y & Y \\
RJ3 &  &  &  &   &  &  &  &  &  & N & --- & ---        & 0.4 & Y & Y & Y  \\
RJ4 &  &  &  &   &  &  &  &  &  & Y & 1   & $10^{13}$  & 0.4 & Y & Y & Y \\
\hline
RS1 & \multirow{4}{*}{$6\times10^6$} & \multirow{4}{*}{100} & \multirow{4}{*}{$2.6\times 10^4$} &\multirow{4}{*}{659} & \multirow{4}{*}{$3.16$} 
& \multirow{4}{*}{0.1}& \multirow{4}{*}{0.08} & \multirow{4}{*}{0.06} & \multirow{4}{*}{5.2} & Y & 12.6 & $ 3\times10^{11}$ & 4.7  & N & N & N(?) \\
RS2 &  &  & & &  & &  & & & Y & 1 & $ 10^{13}$     & 0.6   & N & Y & N \\
RS3 &  &  & & &  & &  & & & Y & 3 & $10^{12}$     & 0.6   & Y  & Y & Y \\
RS4 &  &  & & &  & &  & &  & Y & 1 & $ 10^{13}$     & 0.6   & Y & Y & Y \\
\hline
\end{tabular}
\end{center}
$^1$ whether or not Ohmic dissipation was included. 
$^2$ whether or not the rotation-supported disk formed. 
$^3$ whether or not the protostellar outflow appeared. 
\end{table}


\clearpage

\clearpage

\clearpage

\clearpage

\end{document}